# Mechanisms of Spatiotemporal Mode-Locking

Logan G. Wright[1], Pavel Sidorenko[1], Hamed Pourbeyram[1], Zachary M. Ziegler[1], Andrei Isichenko[1], Boris A. Malomed[2,3], Curtis R. Menyuk[4], Demetrios N. Christodoulides[5], and Frank W. Wise[1]

1. School of Applied and Engineering Physics, Cornell University, Ithaca, NY 14853, USA
2. Department of Physical Electronics, School of Electrical Engineering, Faculty of Engineering, and the Center for Light-Matter Interaction, Tel Aviv University, 69978 Tel Aviv, Israel
3. ITMO University, St. Petersburg 197101, Russia
4. Department of Computer Science and Electrical Engineering, University of Maryland Baltimore County, Baltimore, Maryland 21250, USA
5. CREOL/College of Optics and Photonics, University of Central Florida, Orlando, Florida 32816, USA

**Abstract**
Mode-locking is a process in which different modes of an optical resonator establish, through nonlinear interactions, stable synchronization. This self-organization underlies light sources that enable many modern scientific applications, such as ultrafast and high-field optics and frequency combs. Despite this, mode-locking has almost exclusively referred to self-organization of light in a single dimension - time. Here we present a theoretical approach, attractor dissection, for understanding three-dimensional (3D) spatiotemporal mode-locking (STML).  The key idea is to find, for each distinct type of 3D pulse, a specific, minimal reduced model, and thus to identify the important intracavity effects responsible for its formation and stability. An intuition for the results follows from the "minimum loss principle," the idea that a laser strives to find the configuration of intracavity light that minimizes loss (maximizes gain extraction). Through this approach, we identify and explain several distinct forms of STML. These novel phases of coherent laser light have no analogues in 1D and are supported by experimental measurements of the three-dimensional field, revealing STML states comprising more than $10^7$ cavity modes. Our results should facilitate the discovery and understanding of new higher-dimensional forms of coherent light which, in turn, may enable new applications.

**Introduction**
Light fields created by passive mode-locking (ML) are unmatched for ultrashort duration, peak electric field strength, energy localization, low noise, and frequency precision. These unique features make ML lasers essential for frontier science and technology in physics, biology, chemistry and materials science[1–4] . Today, practically-viable implementations based on fiber or semiconductor gain media hold promise to bring these frontier-pushing technologies to mainstream use[3,5,6], while the state-of-the-art capabilities continue their remarkable growth[7]. Mode-locking in lasers was demonstrated shortly after the laser itself [8] and understood in the frequency domain as the synchronization of longitudinal modes. Longitudinal modes are patterns of the electromagnetic field structured along the length of the resonator, whose resonant lasing frequencies differ by a free spectral range, $\Delta_p = c/P$, where $P$ is the optical path length of one round trip of the resonator. ML occurs when the laser oscillates simultaneously in multiple modes, and the phase difference between them is stable. Of course, laser resonators may support many transverse as well as longitudinal modes. The optical path length for each transverse mode is different, so their free spectral range is different too (Fig. S1b). It was realized early on that if several transverse modes have the right relationship between resonances, multi-transverse-mode ML may occur. This was dubbed "transverse ML" for the case involving a single longitudinal mode, and "total mode-locking" when multiple modes of each kind were involved[8,9].

Why does ML occur? A now widespread intuition is that it represents the laser's solution to an optimization problem defined by the cavity. When lasing is initiated, field fluctuations present in all modes are amplified simultaneously. All complex combinations of the laser's modes compete for the limited energy available from the gain medium. In passive ML, a saturable absorber (an element with loss inversely related to the optical power) is added to the laser cavity. Then, the minimum-loss state becomes one in which many longitudinal modes lase simultaneously, in-phase, as this corresponds to a short pulse in time, with maximum peak power. Since it experiences the least loss each round trip, this lowest-loss configuration is preferentially amplified, eventually dominating all other possible mode configurations.

As the study of mode-locking matured, Ginzburg-Landau equations[10–12] were adopted in recognition of the importance of phase modulations in addition to gain and loss. These models reveal the mode-locked pulse to be a (dissipative) soliton, most commonly when the anomalous group velocity dispersion of the resonator cancels the nonlinear phase shift arising from the Kerr nonlinearity in the laser medium. Later, it was realized that mode-locking also occurs in lasers with normal dispersion[12]. In this case, the pulses carry a linear chirp, and a spectral filter is used to periodically shorten the chirped pulse by attenuating edges of the spectrum (and, owing to the chirp, wings of the pulse in time).

Some works have considered aspects of higher dimensions in ML (e.g., Refs.[13–16]), notably in the Kerr-lens realization, where weak coupling to higher-order modes is essential[17–21]. These works still view ML as a low-dimensional phenomenon. A notable exception is work on random lasers[22–24], but to date coherent self-organization in these systems has been limited. Finally, early studies of multi-dimensional ML [8,9] predated insights from nonlinear dynamics, and so represent only a minute fraction of the possibilities. We recently reported observations of spatiotemporal ML[25] in the form of three-dimensional coherent pulses in lasers based on multimode (MM) fibers. While this work showed self-organized pulses in 3D lasers, physical understanding was limited. Investigation of STML presents acute theoretical and experimental challenges. Direct simulation can be significantly complicated by the presence of disorder, spatiotemporal dispersion and multi-scale nonlinear interactions among a population of modes $10$-$10^4$ times larger than in traditional ML. Similar challenges occur for applying traditional 1D measurement techniques to STML oscillators, as virtually all the standard experimental techniques implicitly assume a spatially single-mode (i.e., 1D) field. Thus, understanding of STML has been restricted to observational reports with careful, but limited, adaptations of traditional theoretical and experimental methods.

We outline here a mechanistic description of spatiotemporal mode-locking (Fig. 1) based on a new theoretical approach that we call attractor dissection. We construct simplified models for 3D mode-locked lasers to identify the key mechanisms for several novel 3D laser solitary waves. This approach is adapted from workhorse techniques of laser physics[10,26–28] and its results inform a similar intuition regarding the laser's steady state. That is, STML steady-states can be intuited as the result of a multidimensional optimization of the laser field toward the eigenstate of the cavity that most effectively extracts energy from the gain medium. To address the experimental challenges, we utilize a flexible approach for 3D electric field measurement, scanning off-axis digital holography, and perform experimental modal decomposition.

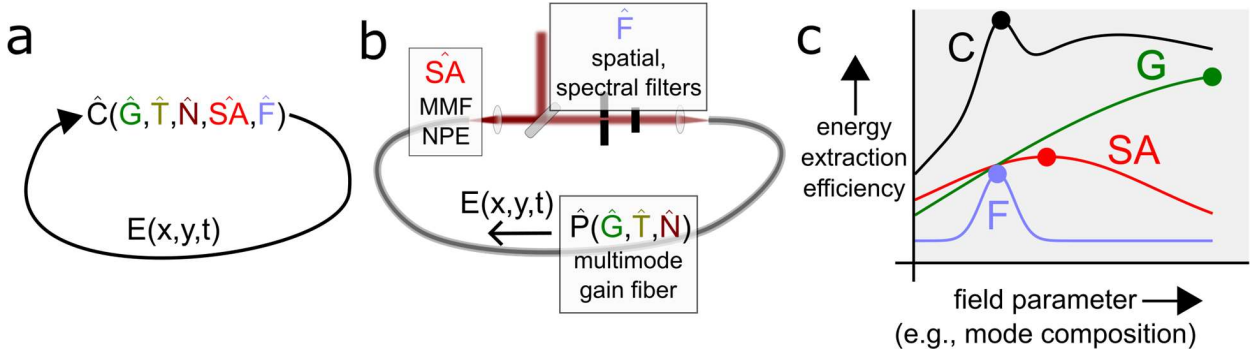

**Figure 1: Conceptual outline of spatiotemporal mode-locking, attractor dissection, and the spatiotemporal maximum-gain principle. a** A laser cavity can be described by a nonlinear projection operator $\hat{C}$, which represents the entirety of the effects of propagation in the course of one round trip through the cavity on the intracavity electric field (Eqn. 1). **b** illustrates how the different effects that compose $\hat{C}$ manifest in our experiments, and our numerical model of, a multimode fiber laser. Here, $\widehat{SA}, \hat{F}$, and $\hat{P}$ are operators describing the effects of the spatiotemporal saturable absorber, intracavity spatial filter and nonlinear gain fiber ($\hat{P}$ includes the inseparable effects of 3D gain, $\hat{G}$, dispersion, $\hat{T}$, and nonlinear mode coupling, $\hat{N}$). **c** illustrates an approximate intuition for the laser's route to steady-state, an *energy extraction efficiency* surface (EEE) **C**, which is the product of the EEE curves arising from the component effects shown, with respect to an arbitrary field parameter (e.g., the amplitude in a given 3D mode). EEE describes how much energy a field with given parameters extracts from the gain medium. A (spatiotemporally) mode-locked pulse $A(x, y, t)$ corresponds to an *eigenpulse* of $\hat{C}$: $A(x, y, t) = \hat{C}A(x, y, t)$, and is an attractive fixed point of the iterative map, i.e., $A(x, y, t) = \lim_{n\to\infty} \hat{C}^n E(x, y, t)$. The eigenpulse typically maximizes EEE. Through attractor dissection, we identify the laser gain competition as an optimization approximately factored into sub-problems. The example displayed in c illustrates a spatial-filter-dominated regime, where the coordinates of the expected steady-state solution (the maximum of **C**, black circle**)** are closest to attractor of F, the spatial filter. In other regimes, the gain (G) and saturable absorber (SA) play more important roles.

**Attractor Dissection Theory**
We now outline the attractor dissection theory and apply it to understand the distinct regimes of steady-state 3D pulse formation that occur as we vary one parameter of the cavity, the intracavity spatial filter size. For further details, see the Methods and Supplementary Sections 1-4. We consider the laser cavity as a nonlinear projection operator representing the transformation of the laser field envelope, $E_i(x, y, t)$, through its propagation along the longitudinal coordinate $z$ of the cavity through one round trip (Fig. 1a). The cavity operator $\hat{C}$ performs the mapping
$$E_{i+1}(x, y, t) = \hat{C} E_i(x, y, t) \qquad (1)$$
where the subscript of $E$ is the round-trip number. $\hat{C}$ accounts for several different effects which are, in general, coupled inseparably. In this view, the laser physics is a composition of iterated nonlinear projection operations. In each round trip, $\hat{C}$'s nonlinear dissipation selects from the field certain attributes, and the saturable laser gain provides a conditional (frequency and energy-limited, and spatially-localized) rescaling of the selected field.

We first consider a simplified multimode laser cavity containing 7 transverse modes, consisting of a normal-dispersion multimode gain fiber, an idealized spatiotemporal saturable absorber (a saturable absorber sensitive to the local field intensity in space-time, $|E(x, y, t)|^2$), a spectral filter, and a spatial filter (Fig. 1b). The fiber is a model of the one used in experiments, whose key feature is its relatively low modal dispersion. As the filtering, saturable absorber and fiber propagation occur in spatially-separate regions, $\hat{C}$ can be factored into $\hat{C} = \hat{F}(x,y)\hat{F}(\omega)\widehat{SA}(x,y,t)\hat{P}(x,y,t)$, where $\hat{F}(x,y)$ and $\hat{F}(\omega)$ are the spatial and spectral filter functions, $\widehat{SA}(x,y,t)$ is the spatiotemporal saturable absorber transfer function, and $\hat{P}(x,y,t)$ accounts for the effect of the pulse propagation through the 3-dimensional nonlinear gain medium. $\hat{P}$ includes the inseparable effects of 3D gain, $\hat{G}$, linear effects such as spatiotemporal dispersion, $\hat{T}$, and nonlinear mode coupling $\hat{N}$. $\hat{P}$ is implemented by integrating a set of 7 coupled nonlinear PDEs.

Holding other parameters fixed, we simulate this cavity, starting from noise, with varying Gaussian spatial filter sizes (positioned for simplicity on-axis with the fiber). Figure 2 shows the spatial mode composition of the steady-state pulses as a function of the spatial filter width (solid dots). We refer to these stable 3D fields, $A(x,y,t)$, which are fixed points of Eqn. 1, as eigenpulses of the cavity. This terminology emphasizes that they are eigenfunctions of the cavity operator: $\hat{C}A(x,y,t) = A(x,y,t)$ (but are not necessarily solitons, as occur in models that take the continuous-z limit of Eqn. 1).

The key idea of attractor dissection is that, despite the many effects occurring each round trip, a given eigenpulse may be principally selected by far fewer. This allows us to construct minimal models of the laser in which $\hat{C} = \hat{R}\hat{O}$, where $\hat{O}$ describes the effect(s) relevant to the specific eigenpulse's attractor, and $\hat{R}$ is a rescaling operator which acts as a simplified saturating gain, returning the field's energy to a fixed value once per round-trip. $\hat{O}$ may be *any* effect or combination of effects in the cavity. By finding for a given eigenpulse the minimal $\hat{O}$ necessary to approximate it, we can identify the key effects responsible for its formation and stability. The fixed point attractor of $\hat{C}$, which we call the field attractor of $\hat{O}$, provides insight into the role of $\hat{O}$ within the full laser dynamics (i.e., what type of eigenpulses does $\hat{O}$ "prefer"?). We can find the field attractors by computing the asymptotic solution to repeated iterations of $\hat{C}$ starting from a noise field, $E_0(x,y,t)$:

$$A(x,y,t) = \lim_{n\to\infty} [\hat{R}\hat{O}]^n E_0(x,y,t). \qquad (2)$$

As an example, consider the field attractor of the spatial filter. If we take $\hat{O} = \hat{F}$ in Eqn. 2 and solve with the 7 modes of the cavity described above, the solution $A(x,y,t)$ has a straightforward interpretation. In the basis of the guided transverse modes of the fiber cavity, $\hat{F}$ takes the form of a matrix $F$ with elements $F_{mn} = \langle \varphi_m | F(x,y) | \varphi_n \rangle$, where $F(x,y)$ is the transmission function of the filter, and $\varphi_i = \varphi_i(x,y)$ are the transverse modes. The matrix $F$ has eigenvectors $\vec{v}_i$, which are complex mode coefficients for fields that pass through the filter with only a uniform loss, i.e. $\hat{F}\vec{v}_i = t_i \vec{v}_i$, where $T_i = |t_i|^2$ is the energy transmission through the filter. The relevant fixed point of Eqn. 2, the attractor of $\hat{F}$, is the lowest-loss eigenvector of the spatial filter in the guided mode basis (with a uniform distribution in time). Curves in Fig. 2 show the modal composition of this spatial filter attractor alongside full numerical simulations from 1 to 9 µm spatial filter sizes. We observe that, even though the dissected attractor model neglects most of the laser physics, it accurately predicts the composition of the eigenpulse in this regime. Thus, we identify the spatial filter as the key mechanism for the eigenpulses in this regime, which can be referred to as *spatial filter-driven pulses*.

We can repeat this procedure for the spatiotemporal saturable absorber, where we use a minimal model $\hat{O} = \widehat{SA}\hat{T}$ (where $\hat{T}$ describes the linear spatiotemporal dispersion of the modes). We find that two distinct fixed point solutions to (2) exist, depending on whether the peak intensity of the pulse is below or above the saturation intensity of the saturable absorber. These dissected attractor models accurately predict the full numerical simulations for filter sizes between about 10 and 30 μm in Fig. 2 (thus, those eigenpulses can be mechanistically described as *saturable-absorber-driven pulses*). Similarly, we find that a minimal model for many of the eigenpulses in the large-filter regime uses $\hat{O} = \widehat{SA}\hat{P}(\hat{T}, \hat{G})$, where $\hat{P}(\hat{T}, \hat{G})$ is the gain fiber propagation operator neglecting nonlinear effects other than the gain, $\hat{G}$.

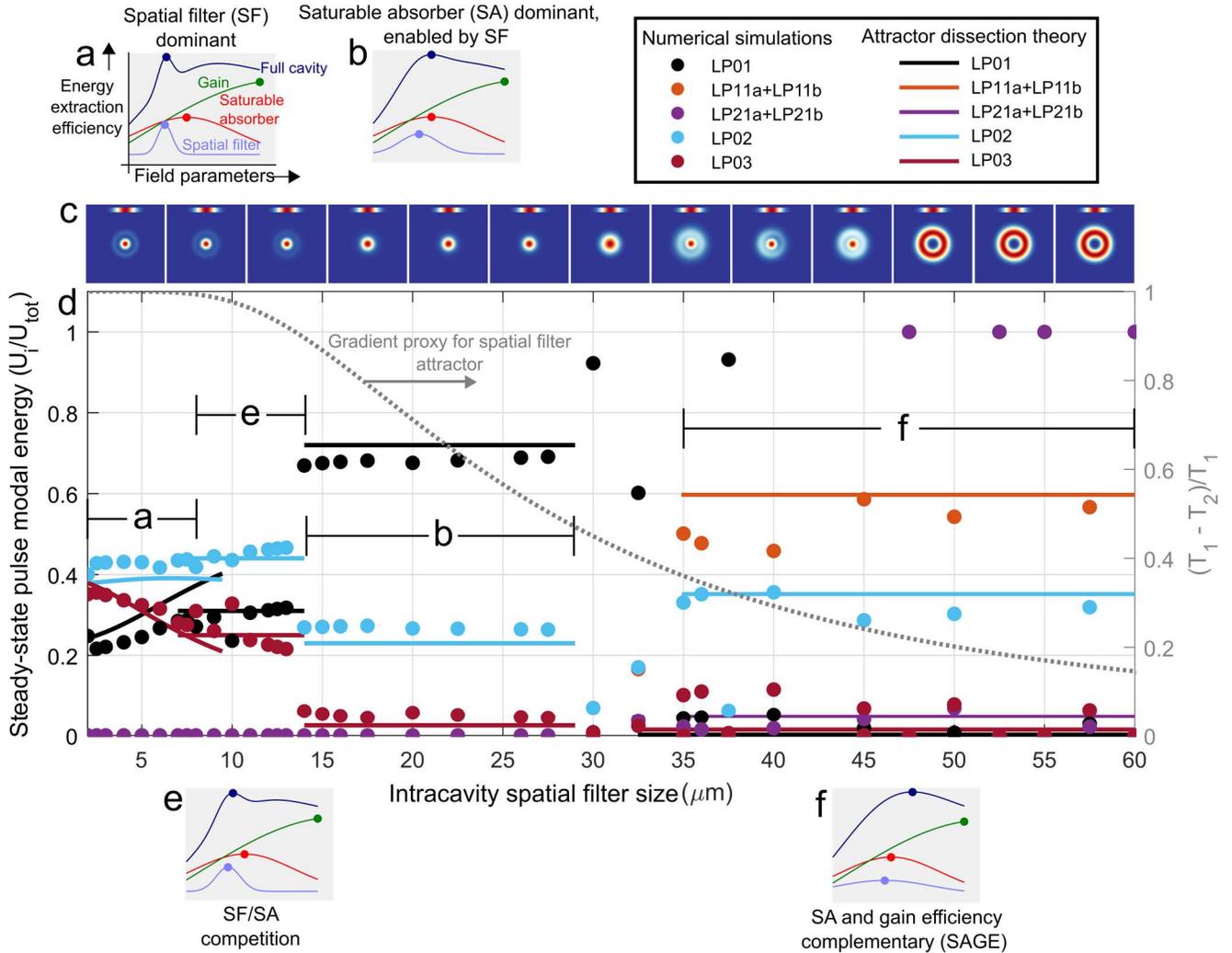

**Figure 2: Identifying the mechanisms of 3D mode-locked pulses for varying intracavity spatial filter size.** The plot in **d** shows the numerically calculated mode composition (solid points: the energy in each transverse mode family relative to the total pulse energy) of the steady-state pulse solution for a multimode fiber-based cavity, for varying size of the spatial filter (SF). The theoretical modal compositions of the relevant attractor models are plotted alongside the numerical points (solid curves). For each regime, representative EEE surface "cartoons" (**a,b,e,f**) are shown to provide intuition for the attractors. The right vertical axis (grey) shows the gradient proxy (see main text) for the spatial-filter's attractor, which estimates the relative importance of the spatial filter in maximizing EEE. As the spatial filter is increased, the steady-state operating regime changes from being dominated by the spatial filter attractor (**a**, SFA) to a saturable-absorber-driven regime enabled by the spatial filter (**b**), to competition between the saturable absorber and SFA (**e**), and finally to a regime characterized by a compromise between minimizing the saturable absorber loss while maximizing the gain efficiency (**f**, SAGE). For reference, in (**c**) representative beam profiles of steady-state pulses are shown above the plot. The scale bar in each plot is the Gaussian profile of the fundamental transverse mode (35 µm MFD). An extended version of the figure and representative details of the steady-state regimes are presented in Supplementary Figures 10-13.

**Maximum gain principle**

To see how the dissected attractor theory mechanistically explain 3D laser physics, a helpful intuition is similar to the minimum-loss principle (this principle is established wisdom in laser science; a useful perspective is in Ref. [29]). In a 3D laser, the expected winner of the gain competition is the field that, through a combination of low loss and high overlap with the gain medium, maximizes the EEE from the gain medium (that is, maximizes the total energy gained by the field).

The idea of an expected winner of the gain competition is an effective intuition for the steady state in a 3D mode-locked laser: the steady state is the laser's attempt to solve an optimization problem, maximizing EEE while satisfying the periodic boundary condition and starting from low-amplitude noise. The path to steady state can be visualized as a maximization on an abstract surface in the 2M-dimensional space of the laser field's 3D mode coefficients, with M the number of transverse modes (Fig. 1c). We can imagine this EEE surface as being composed of distinct contributions from each effect in the cavity, i.e., as a problem approximately factored into distinct sub-problems. Each dissected sub-surface has a maximum that corresponds to that effect's attractor, which is the steady-state field that minimizes (maximizes) the loss (gain) from that effect. The one-dimensional EEE surface cartoons in Figs. 1c and 2 illustrate the intuition: sub-surfaces cooperate or compete in producing the EEE maximum, whose coordinates specify the expected winner of the laser gain competition. In some cases, as with spatial-filter-driven pulses, the maximum of EEE is mostly determined by the maximum of a single sub-problem. In most scenarios, however, multiple effects cooperate or compete to form maxima of EEE.

A quantitative measure of the importance of an effect is the gradient of the EEE surface near its maximum. The solid grey curve (right y-axis) in Fig. 2 shows a proxy measure of the gradient of the EEE surface around the spatial filter's attractor: the normalized difference in transmission between the first and second filter eigenfunctions, $(T_1 - T_2)/T_1$. As the spatial filter expands, the EEE gradient diminishes (from 5 to 10 µm filter widths, visualized in Fig. 2e). Thus, minimizing the loss through the spatial filter becomes less important for maximizing EEE. Eventually, the spatial filter and saturable absorber effects have nearby, similar-height contributions to the EEE surface, so eigenpulses form between the respective attractors. As the filter expands, the saturable absorber becomes more important to the EEE maximization, and the eigenpulses then resemble the saturable absorber's attractor (Fig. 2b). Finally, as the filter is made broader still, eigenpulses change from comprising the few radially-symmetric LP0N modes to a diverse set of higher-order modes. As implied by the EEE surfaces (Fig. 2f), this transition reflects the growing importance of the field's overlap with the 3D gain medium.

We call the eigenpulses in the large-filter regime SAGE (for saturable absorber-gain efficiency) pulses, since in this regime, EEE maximization is approximately factored into (i) maximizing gain efficiency via overlap with the gain medium (accomplished by occupying large-area, high-order modes) and (ii) minimizing saturable absorber loss (accomplished by forming short, intense pulses). This is a multifaceted compromise, and leads to a range of novel 3D physics, which we summarize here (for further details see Supplementary Sections 5-8). Since the gain is slow (temporally nonlocal), EEE maximization can lead initially to numerous distinct 3D pulses forming independently from noise, which independently minimize saturable absorber loss and collectively maximize overlap with the gain medium. These pulses propagate with different group velocities (they contain different modes), so they eventually collide. Through these collisions, single-mode SAGE pulses form as the multimode pulses cross-modulate each other through the spatiotemporally-local saturable absorber (see Supplementary Movie). Intuitively, single-mode pulses are promoted by the saturable absorber due to their minimum modal dispersive broadening. These single-mode SAGE pulses, which are visible in Fig. 2, e.g., at 47.5 and 55 μm filter size in the LP21 mode group, are practically important. They reach higher energy than is possible in single-mode oscillators, are free of spatiotemporal (modal) dispersion, and due to their unique mechanism, should also occur even in step-index fibers (Supplementary Section 8). They are, however, *not* the expected winner of the gain competition: in Fig. 2, these pulses' composition deviates significantly from the expected SAGE attractor. Local-global competition of this kind is one reason that STML oscillators lack the same distinction between single- and multi-pulse regimes as conventional oscillators[30], and single-mode SAGE pulses highlight the limits of the approximate dissected attractor theory and EEE maximization intuition. For a given cavity configuration in the SAGE regime, other qualitatively-different, multistable steady-states emerge as a result of different initial noise conditions. For example, we also observe multimode SAGE pulses that display a periodic breathing. In some cases, they are higher-order eigenpulses, satisfying $A(x, y, t) = \hat{C}^n A(x, y, t)$ for integer $n$, while others display breathing incommensurate with the cavity period due to unconstrained transverse mode beating (see Supplementary Section 7).

**Reduced models: building up complex attractors from dissected attractors**
A motivation for attractor dissection is that it naturally extends to more advanced, but still reduced, models to describe experiments. The simplest models – the ones considered so far – are useful for identifying the primary mechanisms for distinct 3D mode-locking regimes in full numerical simulations with 7 modes. Simulations of fiber that supports 180 modes (as in experiments described below) is challenging, especially because of the complex role of disorder – an experimental feature we find essential for accurate modelling. While these reduced models fail to describe the details of ultrafast pulse propagation, they permit efficient simulations and provide insight into experiments, clarifying how effects like disorder affect the regimes identified in the few-mode, disorder-free models. To compare with experiments, we use a reduced model for $\hat{C} = \hat{R}\hat{T}_\text{r}\hat{F}\widehat{SA}\hat{G}$, where $\hat{G}$ is an approximation of the saturating gain operator, and $\hat{T}_\text{r}$ is the linear transfer function of the cavity, extended to include random linear mode coupling (see Methods, Supplementary Section 4 for details).

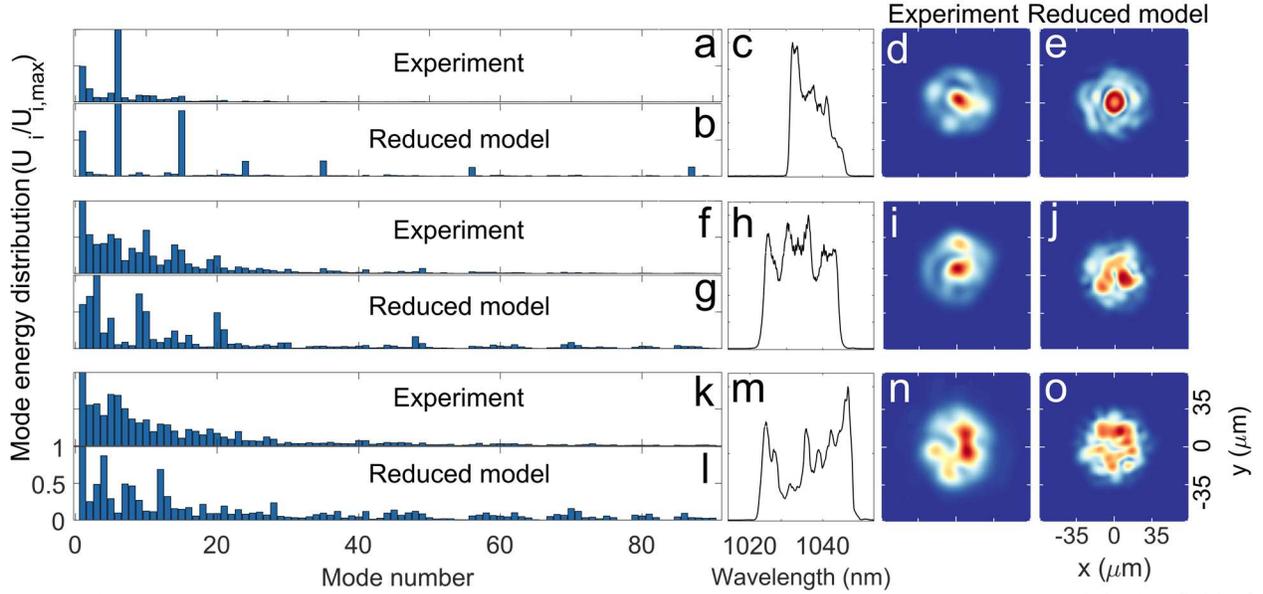

**Figure 3: Experimental regimes of STML and results from a reduced laser model.** We measured the 3D field of stable mode-locked pulses in a normal-dispersion multimode fiber laser for various intracavity spatial filter dimensions. **a-e** show results for a pulse with a narrow spatial filter in the cavity (roughly spatial filter driven pulses), with **a** showing experimental mode decomposition of the 3D field into the transverse fiber modes, **b** the same for a reduced model calculation, **c** the experimental pulse spectrum (note the y-axis is normalized to the mode with the maximum energy), and **d** and **e** the near-field beam profiles of the experimental and reduced model pulses respectively. **f-j** and **k-o** show the same series of plots for mode-locked states with an intermediate-sized intracavity spatial filter, and for a cavity without deliberate spatial filtering (roughly, SAGE pulses) respectively. An extended version showing more typical experimental results is Supplementary Figure 14, while Supplementary Figures 16-17 show additional measurements. Here, the reduced model neglects the Kerr nonlinearity of the fiber and employs approximate operators to efficiently describe the saturating gain and linear fiber disorder. Despite these approximations, the reduced model reproduces coarse trends of experiments, such as the reducing symmetry and broader modal distribution as the filter size is increased.

**Experiments**

Experimentally, we observed STML in lasers constructed with multimode graded-index Yb:doped gain fiber and some free-space sections, similar to Ref. [25]. The laser incorporates an adjustable spatial filter, a spectral filter, and a spatiotemporal saturable absorber, implemented using nonlinear polarization evolution (NPE) in the MM gain fiber. For a suitable choice of waveplate orientations in the cavity, NPE amounts to the action of an effective spatiotemporal saturable absorber, which imposes a loss inversely related to the spatiotemporal intensity of the circulating field. The electric field is measured in three dimensions using scanning off-axis digital holography (see Methods) and is then decomposed into the transverse modes of the fiber (Fig 3).

As we vary the spatial filter experimentally, we observe, as in the few-mode simulations in Fig. 2, various transitions in the STML steady-state pulses (Fig. 3, Supplementary Figure 14). These pulses have similar spectra, pulse duration, energy and other features as those in our simulations. Spatial-filter-driven pulses (narrow spatial filter) are narrowband, with triangular spectra and radially-symmetric beams. With intermediate filter sizes, pulses are similar to saturable-absorber-driven pulses; they have beam profiles with near-radial symmetry, have broad rectangular spectra and concentrate energy in low-order modes. SAGE (broad or no spatial filter) pulses are broadband and energetic, with rectangular spectra and multimode beam profiles. Qualitative dynamical features such as multistability and multiperiodicity are observed; much like the 7-mode simulations, these occur primarily in the large-filter regime.

Beyond these qualitative trends, the experimental results are more complex (Fig. 3, more details in Supplementary Sections 4, 6-7). Reduced models are helpful for predicting how the effects of disorder, and the increased dimension of the optimization, affect the regimes of STML. Both in experiment and reduced models, we see that with narrow spatial filters, pulses still comprise mostly radially-symmetric modes. As the filter is widened, however, disorder and modal diversity incrementally result in more complex pulses. We routinely observe SAGE pulses containing over 30 million locked 3D modes, about 10 times more than in any previous mode-locked oscillator [31,32]. This quantitative comparison fails to capture the qualitative strangeness of these pulses, however. Characterized by disordered spatiotemporal structure, the experimental SAGE pulses nonetheless exhibit remarkable long-range order (evidence by high radio frequency contrast > 80 dB, Supplementary Figures 16-17). They are robust, with many being self-starting and stable over long durations and perturbations in the laboratory. While principal modes of disordered fibers[33,34] suggest a first guess toward a detailed model of these pulses, reconciling the surprising complexity with the full multimode nonlinear dynamics and periodic boundary conditions of the cavity will – like many other new phenomena in 3D lasers – require advances in both theoretical and experimental methods.

**Discussion**

The present study focuses on a single narrow swath of parameter space – the variation of the spatial filter in a normal dispersion, MMF-based STML oscillator. STML oscillators present many possibilities for exploring multidimensional nonlinear dynamics and solitary waves, and fundamentally new forms of coherent light. For example, STML states that approximate MM solitons[37,38] should form with anomalous dispersion, and other spatial and spatiotemporal attractors[13,39,40] should be observable by judicious design of the cavity. Meanwhile, high performance lasers should be possible in settings ranging from few-mode step-index fibers (Supplementary Section 8) to free-space cavities with rod-like fibers, thin-disk, or other bulk gain media. Realizing STML with more systematic intracavity control (e.g., intracavity spatial light modulators or deformable mirrors) and novel spatiotemporal saturable absorbers will be an important step. Advances in 3D measurement tools will be crucial for insightful experimental studies, and, hopefully, automated closed-loop control, of STML lasers[41–44].

While our intention here has been to use the optimization-problem perspective to derive intuitive models for 3D laser physics, future reconfigurable 3D lasers are promising as analogue computers and heuristic optimizers. Even at a modest scale of 180 transverse modes, the cavity's natural computations are ≈10-100 billion times faster (and require ≈$10^{13}$ times less energy) than full numerical simulations on a digital computer. 3D lasers have many more controllable degrees of freedom than 1D oscillators, and have been described approximately with spin-glass models[23,29,45] similar to coherent Ising machines[46]. While the physics of competitive optimization (i.e., Darwinism) is itself a universal model for a wide range of physics, from materials to biological evolution, extending mathematical spin-glass and synchronization analogies will be helpful for using 3D lasers as heuristic optimization and analog simulators. For experimentally realizing these and other applications, control of coupling using active intracavity modulation[47], or by engineered modes in semiconductor lasers[48], multimode microresonators[49] or coupled microresonators[50] are promising.

In this work, we have introduced a conceptual framework that can readily be applied to develop a mechanistic understanding of the complex mode-locking phenomena that occur in 3D lasers. We have shown that this approach is consistent with numerical and analytic models derived in limiting cases and can describe experimentally-observed coherent phases of light in a multimode fiber laser. This enables us to observe and explain new dynamical phenomena and coherent phases of light that have no analogues in lower dimensions. Much like 1D mode-locking eventually served as a uniquely enabling tool for previously-inconceivable measurements and applications, we anticipate that control of self-organized coherent light in higher dimensions using spatiotemporal mode-locking may enable applications that today appear fantastical.

**Methods**
**Theory - Nonlinear wave equation simulations**
The main numerical and theoretical models are outlined in Supplementary Section 2-4. For our few-mode simulations, we consider only 7 modes: the first 6 modes, plus the 15$^{th}$ one (LP03). For simulations presented in the main article, we consider a 50-cm long fiber, which is chosen to be a similar order-of-magnitude as the experiment, but short enough to permit thousands of round trips to be efficiently simulated for each data point in Fig. 2. We take the gain to be a spatially-saturating function of the fluence (the integral of the intensity over the temporal window) at each point in the transverse plane. The spatiotemporal saturable absorber is represented by an idealized transfer function with a modulation depth of 1 and a saturation intensity of 50 GW/cm$^2$. Both numbers are of similar magnitude to those routinely used to describe normal-dispersion single-mode fiber lasers based on nonlinear polarization evolution. Besides the parameters noted, we assume an output coupling ratio of 0.7, an additional lumped loss of 0.5, and a Gaussian spectral filter of 10-nm full-width-at-half-maximum. In addition, we rescale the field after passing through the spatial filter. This artificial step is done to keep the lasing threshold approximately constant as the filter size is varied, and thus allow the spatial filter size to be adjusted without changing the gain saturation energy. Since dissipative projection followed by the energy rescaling of the saturating gain is the essence of the laser physics, this approximation does not change the physics qualitatively.

**Theory – Dissected attractor laser models for component attractors and reduced laser models**
Supplementary Sections 3 and 4 describe the calculation of the attractors of each intracavity effect, and extensions of these approximate laser models to more complex attractors within slightly more-realistic laser models. Like the main theoretical model, these ones describe the laser as action of an iterative nonlinear projection operation on the intracavity field. The key difference is that only one or a small number of specific effects are considered alongside idealized implementations of the laser cavity's physics, linear modal propagation and rescaling (from the saturating gain). In other words, all such models are based on an equation of the form:
$$A(x,y,t) = \lim_{n\to\infty} \hat{C}^n E_0(x,y,t) = \lim_{n\to\infty} [\hat{R}\hat{O}]^n E_0(x,y,t) \qquad (3)$$
where $\hat{O}$ represents an effect (or small number of effects) of interest, $E_0(x,y,t)$ is an initial noise field, and $\hat{R}$ is the unconditional rescaling operation representing an idealized saturating laser gain. $\hat{R}$ amounts to the following:

1. Calculate the total energy of the field, $U = \iiint |E(x,y,t)|^2 dx dy dt$
2. Multiply the field by $\sqrt{U_0/U}$, so that the energy of the field is restored to a constant $U_0$ regardless of the losses or gains elsewhere.

In most of the minimal models, we need the operator $\hat{T}$. $\hat{T}$ is the linear transmission operator that describes linear propagation of the cavity's guided modes (defined by the presence of the fiber, the bounding mirrors, etc.) through one round trip. The linear transmission $\hat{T}$ is accomplished by multiplying the field in the frequency domain, $\tilde{A}_{q\omega}$, by a transmission matrix, $T_{qp\omega}$, for each transverse mode $q$ and each frequency $\omega$,

$$\tilde{A}_{q\omega} \to \sum_{p=1}^{M} \tilde{A}_{q\omega} T_{q,p,\omega} \quad (4)$$

where $M$ is the number of transverse modes. For the calculation of the component attractors relevant to the few-mode simulations, $T_{q,p,\omega}$, is diagonal (since disorder is neglected), so $T_{qp\omega} = T_{q\omega}\delta_{qp}$, is derived as the solution to the linear fiber propagation over the fiber length from $z = 0$ to $z = L$:

$$\tilde{A}_{q\omega} \to \tilde{A}_{q\omega} \exp\left[i\delta\beta_0^{(q)}L - i\Omega\delta\beta_1^{(q)}L - \sum_{m=2}^{3} i^{m+1} \frac{\beta_m^{(q)}}{m!} \Omega^m L\right] \quad (5)$$

where $\Omega = \omega - \omega_0$. In general, $\hat{T}$ may also account for linear coupling between guided modes, such as is caused by bending or clamping a multimode fiber, by imperfections in the imaging from the fiber end facet to end facet, or possibly by coupling deliberately introduced, e.g., by waveplates, two-dimensional phase plates, dispersive delay lines, or other components.

We have typically obtained the solutions to Eqn. 2 numerically, initializing the field as a noise input and performing a large number of iterations until a steady-state is attained.
As mentioned in the article, such models can be naturally extended into a hierarchy of more accurate, but more complex, models of the laser. For example, to model the experiments with all the fiber's guided modes in Fig. 3, we apply this by neglecting details of nonlinear pulse propagation. This reduced model corresponds to an equation of the form

$$A(x,y,t) = \lim_{n\to\infty} \left[\hat{R}\hat{T}_r\hat{F}\widehat{SA}\hat{G}\right]^n E_0(x,y,t), \quad (6)$$

where $\hat{T}_r$ is the linear transmission through the cavity as above, but including disorder. $\hat{R}$ is obviously artificial and is not strictly required in reduced laser models; it is applied here practically to minimize the need to systematically vary the gain saturation energy when other parameters are modified. Here, we use a simplified model of the gain operator by employing an approximate solution to the field propagation (see Supplementary Section 4 for details). Disorder is implemented in $\hat{T}_r$ by multiplying the vector of modal coefficients at each frequency before and after applying the spatiotemporal dispersion by a matrix $M_D = I + rX$, where $I$ is the $M$ by $M$ identity matrix, $r$ is a parameter characterizing the strength of disorder (between 0 and 1) and $X$ is a random complex matrix (drawn from a uniform distribution from -1 to 1 for real and imaginary components). As experimental disorder manifests both in conservative (coupling between guided modes) and dissipative coupling (i.e., disordered mode-dependent loss through disorder-induced coupling to radiating modes), $X$ is not symmetric. We implicitly assume disorder only couples transverse modes, and take the disorder to be fixed across the timescale of the simulation. Thus, we apply two different, fixed matrices $M_D^{(1)}$ and $M_D^{(2)}$ before and after applying the spatiotemporal dispersion each round trip ($\hat{T}$ as in Eqn. 5). For the simulations shown in Fig. 3, we additionally take the output field to be the rejected light from the saturable absorber, as in the experiment.

For the reduced model simulations in Fig. 3, for the narrow spatial filter simulation we used a 6 µm FWHM Gaussian spatial filter, offset from the center by 1 µm, a 40 µm gain dopant width parameter (see Supplementary Section 4 for definition), $r = 0.05$, and a saturable absorber saturation intensity of 40 GW/cm$^2$, and gain saturation fluence of 3 mJ/cm$^2$. For the intermediate filter simulation, these parameters were respectively 50 µm, 2.5 µm, 35 µm, 0.15, 30 GW/cm$^2$, and 3 mJ/cm$^2$. Finally, for the broad filter simulation, the parameters were 1000 µm spatial filter FWHM, offset by 270 µm, with a gain dopant width parameter of 35 µm), $r = 0.2$, saturable absorber saturation intensity 25 GW/cm$^2$, and gain saturation fluence 10 mJ/cm$^2$.

**Experiments – STML oscillator**
The STML oscillator is based on the same partially-graded MM gain fiber used in Ref.[25]. The laser design is shown in Fig. 1b; more detailed schematics are given in Supplementary Figure 18, and a brief tour of the oscillator and measurements is shown in the Supplementary Movie. A 10-nm spectral filter is accomplished through a birefringent filter, a quartz plate between the output polarizing beam splitter (which also serves as the rejection port for nonlinear polarization rotation) and the isolator. An adjustable spatial filter is realized with an adjustable iris on a 3-axis translation stage. A 60-W, 976-nm pump diode is coupled into the outer cladding of the fiber and combined/separated with the laser signal, using dichroic mirrors. The results presented here are based on a cavity design where the spatial filter is placed in the near-field image of the output fiber end facet, which is subsequently imaged back onto the input fiber end facet, with a 1:1 total magnification. The spatial filter is usually slightly defocused from the near-field image of the output fiber end facet, to achieve a slightly smoother spatial filter with our hard aperture. However, details like this, or even the imaging system itself, are mentioned for completeness only: we have obtained qualitatively similar results in a variety of other designs, including ones with different spectral filter sizes, spatial filter implementations, without imaging optics in the free-space section, and with spatial filters in the far-field plane, or intermediate locations.
We either angle-cleave the fiber at each end or employ angled end caps in order to prevent back-reflections that disrupt mode-locking. While end caps increase the laser's peak power handling and dramatically reduce the effects of clamp-induced linear mode coupling and disordered mode-dependent loss, we found evidence for beam distortions due to the relatively small aperture of our end caps (400 µm). To avoid the error this would introduce in mode decomposition, the results presented here use simply angle-cleaved fibers. We expect that use of larger end caps in the future will alleviate several practical challenges with the current design.

The laser is aligned to maximize the continuous-wave output power, and the spatial filter is aligned to the center of the beam by closing it incrementally and maximizing the output power by adjusting the transverse position. The spatial filter is then opened, and the pump power is increased to a level where mode-locking is expected (in our experiments, this is >20 W). To mode-lock the laser, we employ the same techniques used to mode-lock single-mode fiber lasers based on nonlinear polarization saturable absorbers, but also adjust the spatial filter. Generally, it is easy to achieve mode-locking, but in some realizations finding a stable mode-locked state (i.e., one that remains mode-locked for many hours, and through environmental perturbations over the course of a day) can be challenging. Empirically, we find that the use of a moderate or narrow spatial filter is helpful; this may result from forcing the laser to mostly occupy low-modal-dispersion, low-order modes, and also by reducing the number of competing solutions. A high-quality, self-starting mode-locked state, stable enough to make measurements over several days, can be typically obtained by finding any mode-locked state, then slightly adjusting the pump power and spatial filter size/position to ensure the single-pulse operation and/or stability of the pulse train.

For each mode-locked state, to verify single-pulsing operation without continuous-wave background, we make the following measurements. First, the laser's autocorrelation is measured using a long-range intensity autocorrelator, and the pulse train from the laser is measured with a fast photodiode and oscilloscope. Together, these measurements ensure that only a single pulse exists in the cavity. Note that, due to the spatiotemporal structure of the pulse, the intensity autocorrelation is not an accurate representation of the pulse but is nonetheless useful as a rough measure of the pulses' existence and temporal extent. Second, after spatially-uniform sampling by glass windows, we couple the entire field into a long highly-multimode step-index fiber, and then measure the spectrum using an optical spectrum analyzer. This is done to average the spectrum across the entire field in space, to ensure that no continuous wave background is present (given the numerous lasing modes, it is possible for the continuous-wave background to exist in modes uninvolved with the mode-locked state, hence measurements that do not average across the entire field may yield misleading results). We then measure the radio frequency spectrum using an RF spectrum analyzer, with 10 Hz resolution bandwidth, and 1 Hz video bandwidth. Finally, 3D measurements and mode decomposition are measured as described in the next section. Before and after measurements, we block the cavity and unblock it to allow the laser to start from noise, in order to determine whether the state is self-starting. We find that stable states are almost always self-starting (those reported here are), and are robust against small perturbations of the fiber, the cavity alignment, etc. Some states (especially in the SAGE regime) are bi- or multi-stable ones, hence each time the laser is started it reaches one of several different steady-states. Supplementary Figures 16 to 17 show the complete set of measurements for two representative STML states (both states were additionally verified to be self-starting).
In experiments, the rejected light from the NPE saturable absorber is used as the output coupled light (Supplementary Figure 18). This means that the output field differs slightly from that which remains circulating in the cavity, unlike what takes place in the simulations. To account for this, we have performed some experiments with a second additional output, released after the polarizing beam-splitter that realizes loss for the NPE saturable absorber. With this measure, no significant differences are observed, therefore for practical purposes we maintained the simpler design of using the rejected light as the output coupled light.

**Experiments – 3D pulse field measurement and mode decomposition.**
Our approach to measuring the 3D electric field (3D pulse) emitted by the laser was inspired by work on STRIPED FISH[51,52] and TERMITES[53], as well as by a large number of works on algorithm-based mode decomposition (e.g.,[54–56]). Many other works have been published on both 3D field measurement and mode decomposition, each with their own advantages and disadvantages. The primary objective of our approach was to build a simple device that would have an easily-adjustable space-time-bandwidth product to allow measurement of the range of possibly complex chirped pulses emitted directly by the laser. This was accomplished by the use of delay-scanned off-axis digital holography[57,58] (for a schematic of the device, see Supplementary Figure 18) where a spatially-filtered version of the main pulse was used as a spatiotemporal reference. We measured the electric field $E(x, y, \tau)$, and performed a Fourier transform to obtain $\widetilde{E}(x, y, \omega)$. The phase relationship between frequency components is obtained by measuring the FROG trace of the spatially-filtered reference field. To perform the mode decomposition, we then take the overlap of the field at each frequency with the calculated spatial modes, $\varphi_n(x, y)$, for the nominal refractive index profile of the multimode gain fiber, i.e. $c_n(\omega) = \langle \varphi_n | \widetilde{E}(x, y, \omega) | \varphi_n \rangle$. The plotted values in Figs. 3, $\frac{U_i}{U_{\text{tot}}}$, are then

$$\frac{U_i}{U_{\text{tot}}} = \frac{\int c_i(\omega) d\omega}{\sum_{n=1}^{N} \int c_n(\omega) d\omega} \quad (7)$$

The mode decomposition has a notably high uncertainty following, first, from inexact knowledge of the fiber index profile, and, second, from inability to locate the center of the fiber in the image with high precision. For a fiber with 90 modes, the representational capacity of the guided modes is high, and so the decomposition error associated with the latter effect can be considerable, even if one takes a rigorous approach to image calibration; note that a high-quality mode decomposition can be obtained even when these calibration parameters are wrong. (In contrast, it is much easier to unambiguously obtain the correct calibration for a few-mode fiber, with its very limited representational capacity). Our calibration approach involved utilizing 4 independent, distinct calibration methods of the scale and center position of the fiber, then averaging across these. To translate the uncertainty in the calibration of the scale and fiber center position into the mode decomposition error, we compare the mode decomposition obtained with the nominal optimum calibration versus one where the scale and fiber center position are adjusted by the error of these calibration parameters (inferred by the deviation across the independent calibration methods). In the Supplementary Material, we analyze this error for two specific STML states for which all measurements are presented. We find a typical error of about 6% (8% and 4% for the two examples considered). However, we note that this error estimate is an average across the modes, and specific modes have larger uncertainty (see Supplementary Figure 19), and furthermore it does not account for the error associated with the real fiber's refractive index profile differing our best estimate (which introduces additional error due to the difference between the decomposition modes and the true modes of the fiber). Altogether, these issues motivate our interpretation of the mode decomposition as a rough estimate: it yields a good measure of the overall distribution of energy (in low vs high-order modes, say), but not a precise measurement of the exact modal distribution.

**Acknowledgements**
Portions of this work were supported by the Office of Naval Research (N00014-13-1-0649 and N00014-16-1-3027) and the National Science Foundation (ECCS-1609129, ECCS-1912742). LGW acknowledges helpful discussions with Alexander Cerjan and Hiro Onodera.

# Supplementary Material for

# Mechanisms of Spatiotemporal Mode-Locking

**Logan G. Wright[1], Pavel Sidorenko[1], Hamed Pourbeyram[1], Zachary M. Ziegler[1], Andrei Isichenko[1], Boris A. Malomed[2,3], Curtis R. Menyuk[4], Demetrios N. Christodoulides[5], and Frank W. Wise[1]**

1. School of Applied and Engineering Physics, Cornell University, Ithaca, NY 14853, USA
2. Department of Physical Electronics, School of Electrical Engineering, Faculty of Engineering, and the Center for Light-Matter Interaction, Tel Aviv University, 69978 Tel Aviv, Israel
3. ITMO University, St. Petersburg 197101, Russia
4. Department of Computer Science and Electrical Engineering, University of Maryland Baltimore County, Baltimore, Maryland 21250, USA
5. CREOL/College of Optics and Photonics, University of Central Florida, Orlando, Florida 32816, USA

The supplementary material in this document is organized as follows.

**Section 1** provides a description of spatiotemporal mode-locking (STML) in the frequency domain, in terms of the resonant frequencies of the cavity's modes.

**Section 2** details our primary numerical models, based on propagation of the laser field through the gain medium using generalized nonlinear Schrödinger equations (NLSEs), plus additional effects. These models include most relevant effects, but our use of them mainly focuses on the computationally-efficient case where only a small number of transverse modes are considered. These models are generalizations of the most widely used models for describing ultrafast lasers and nonlinear pulse propagation in the modern literature.

**Section 3** details the treatment of individual effects within the cavity in the simplified description of STML outlined in the paper. Specifically, we develop the nonlinear projection operations for each relevant effect and show the calculation of the attractors for that effect through Eqn. 2 in the main article.

**Section 4** describes reduced models which incorporate a subset of the effects considered in the primary numerical models, by combining components described in Section 3. Extending the approach from Section 3, these models are quite approximate as whole-laser simulations but are computationally compact enough to be applied to conditions relevant to our experiments in a 90-transverse mode fiber, and simple enough to be interpreted easily. Accordingly, they help bridge the gap between experiments and the nonlinear wave physics of STML studied primarily in the few-mode case.

**Section 5** summarizes relevant findings from the models described in Section 2 to illustrate how the different effects within the STML laser interact within the steady-state pulse's nonlinear evolution through the cavity. This section's primary purpose is to illustrate how, in the simplified case of a small number of transverse modes, the different steady-state STML regimes highlighted in the main article correspond to distinct self-consistent nonlinear pulse evolutions.

**Section 6** contains additional supplementary figures mentioned in the main article, particularly regarding the experimental results.

**Section 7** summarizes types of multiperiodic states observed, provides hypotheses for their integration into the theoretical framework considered in the main article, and gives an experimental example.

**Section 8** summarizes design guidelines for STML oscillators, the design of the few-mode step-index fiber for STML and presents the conclusions of a numerical study to develop design guidelines for high energy few-mode fiber oscillators.

**Section 9** contains references for this document.

# Table of Contents

**[Supplementary Section 1: Description of STML in the frequency domain]**

Mode-locking is often considered in the frequency domain. STML is easily reconciled with this picture as a straightforward generalization. In this section, we provide an introductory level description of STML in the frequency domain.

In mathematical terms, the output of a laser (or at any point inside the resonator) can be written in terms of its 3D modes, $M_n(x,y,z)$, and their resonant/oscillation frequencies, $\omega_n$:

$$E_{\text{laser}}(x,y,z,t) = \sum_n c_n(t) M_n(x,y,z) e^{-i\omega_n t} \qquad (S1)$$

When more than one $c_n$ is non-zero in Eqn. S1, and the relationship between the non-zero $c_n$ is stable over many round trips, the laser is spatiotemporally mode-locked.

This description in terms of the modes ignores multiple nuances of the nature of 'modes' within a non-Hermitian system such as a laser or externally-driven lossy resonator [1]. However, as has been done for low-dimensional mode-locking historically, ignoring these nuances is suitable for the high-level description that is our goal here. Often, the 3D modes in Eqn. S1 may be separated into transverse and longitudinal components, $M_{pq}(x,y,z) = L_p(z)\varphi_q(x,y)$. In most mode-locked light sources, the only available transverse mode is the fundamental mode of the resonator, and so $M_n(x,y,z) = L_n(z)\varphi_1(x,y)$, and in essence the laser dynamics are 1+1-dimensional (one space dimension plus time).

To understand STML in terms of mode resonances, in the frequency domain, we first find the mode resonances of a typical multimode laser cavity. To do this, it is easiest to take the assumption that the modes can be factored as $M_{pq}(x,y,z) = L_p(z)\varphi_q(x,y)$. The resonant frequency of the mode with transverse mode index $q$ and longitudinal mode index $p$, $\omega_{pq}$, is found by the condition that the optical round trip length for that mode is an integer number of wavelengths. The optical round trip length for a given mode is $P_{pq}(\omega_{pq}) = n_{pq}(\omega_{pq})L$, where $L$ is the cavity round trip length and $n_{pq}(\omega_{pq})$ is the refractive index of the mode with indices $q$ and $p$, at its resonance frequency. That is, the angular resonant frequency (in rad/s) is:

$$\omega_{pq} = \frac{2\pi c\, p}{n_{pq}(\omega_{pq})L} \qquad (S2)$$

In the case of a multimode fiber laser, if we neglect coupling between transverse modes due to imperfections or perturbations (e.g., bending) in the fiber, or in a free-space section between ends of the fiber, then $n_{pq}(\omega_{pq})$ is the effective refractive index of the fiber's transverse mode $q$ at frequency $\omega_{pq}$. For other resonators, the optical round trip length will need to be supplemented with geometric factors such as the Gouy phase shifts, or other sources of coupling between the waveguide's modes might need to be considered.

By substituting $n_{pq}(\omega_{pq}) = \beta_{pq}(\omega_{pq})/\omega_{pq}c$ into Eqn. S2 and expanding the modal propagation constant $\beta_{pq}(\omega_{pq})$ in a Taylor series around some reference frequency $\omega_0$, one finds (up to 2nd order):

$$\omega_{pq} = \omega_0 + \Omega_p \frac{\beta_1^1}{\beta_1^q} - \Omega_p^2 \frac{\beta_2^q (\beta_1^1)^2}{2(\beta_1^q)^3} \qquad (S3)$$

Here, $\Omega_p = 2\pi(p - p_0)/L\beta_1^1$ is the ideal (without dispersion) angular free spectral range for the fundamental transverse mode family, and we have written the $n$th order term in the Taylor expansion of the $m$th modal propagation constant as $\beta_n^m$. We have also simplified Eqn. S3 by expanding all modes about a common reference frequency $\omega_0$ rather than considering the variation across transverse mode families.

Eqn. S3 shows that the local free-spectral range, $\Delta_q(\omega_{pq}) \approx \frac{1}{2\pi} \frac{\partial \omega_{pq}}{\partial p}$, varies with frequency linearly according to the group velocity dispersion of that mode family, $\beta_2^q$ (Fig. S1a). When mode-locking occurs in a single transverse mode family, $q$, the last term in Eqn. S3 is cancelled by nonlinear, linear, and dissipative interactions between the modes. As a result, the resonant frequencies form a regular frequency comb of narrow resonances evenly spaced by a constant free-spectral range $\frac{\Omega_p}{2\pi} \frac{\beta_1^1}{\beta_1^q}$. This free spectral range differs between different transverse mode families as a consequence of the transverse modal dispersion (Fig. S1b).

It is also useful to consider the phase shift of each mode. For light propagating linearly within the resonator, the phase shift per round trip of each modal component is

$$-\frac{d\varphi_{pq}}{d\tau_{pq}} \approx \frac{\beta_{pq}(\omega_{pq})L}{\frac{L}{v_q}} = \beta_{pq}(\omega_{pq}) v_q \approx \omega_o + \Omega_p \frac{\beta_1^1}{\beta_1^q} + \Omega_p^2 \frac{\beta_2^q (\beta_1^1)^2}{2(\beta_1^q)^3} \qquad (S4)$$

where $\tau_{pq}$ is the time for a single round trip in the mode with indices $q$ and $p$, and $v_q = 1/\beta_1^q$ is the group velocity (at the reference frequency) of the transverse mode group $q$.

When the laser is mode-locked in the fundamental mode, such as in any conventional mode-locked oscillator, its spectrum is a precisely equally-spaced comb, $\omega_p = \omega_o + \Omega_p$. Eqn. S4 shows how, when the final term in Eqn. S3 is cancelled and mode-locking is established, the phase shifts of the field's modal components vary only linearly in frequency (since the final term in Eqn. S4 is also cancelled out). Equivalently, the laser field has a constant group velocity, $v_1 = 1/\beta_1^1$. In the frequency domain, the mode-locked laser field is characterized by narrow peaks, each in phase. The Fourier transform of such a comb of in-phase spectral peaks is a train of regular pulses in time. This regular train is of course just the result of the circulating eigenpulse of the cavity (which is coupled out once per round trip).

When the laser is spatiotemporally mode-locked, everything in the previous paragraph remains true (Fig. S1d). The only additional consideration, besides the presence of multiple transverse modes, is that the group velocity of the mode-locked pulse is not, in general, only given by any transverse mode's group velocity at $\omega_o$, $v_q = 1/\beta_1^q$. In a laser with a single transverse mode group, the mode-locked pulses' group velocity is intermediate between the linear group velocities of its component longitudinal modes. Likewise, in general when spatiotemporal mode-locking occurs the group velocity of the STML pulse (the free-spectral range of the STML pulse spectrum) is intermediate between the linear group velocities (local free spectral ranges) of its component longitudinal and transverse modes.

Higher-order eigenpulses of the cavity operator (Supplementary Section 7) correspond to subharmonic spatiotemporally mode-locked states, which have a super FSR equal to the integer of the subharmonic relative to a typical STML state (Fig. S1e). Other multi-periodic coherent mode-locked states may correspond to similar spectra, except without a super-FSR that can be expressed as an integer multiple.

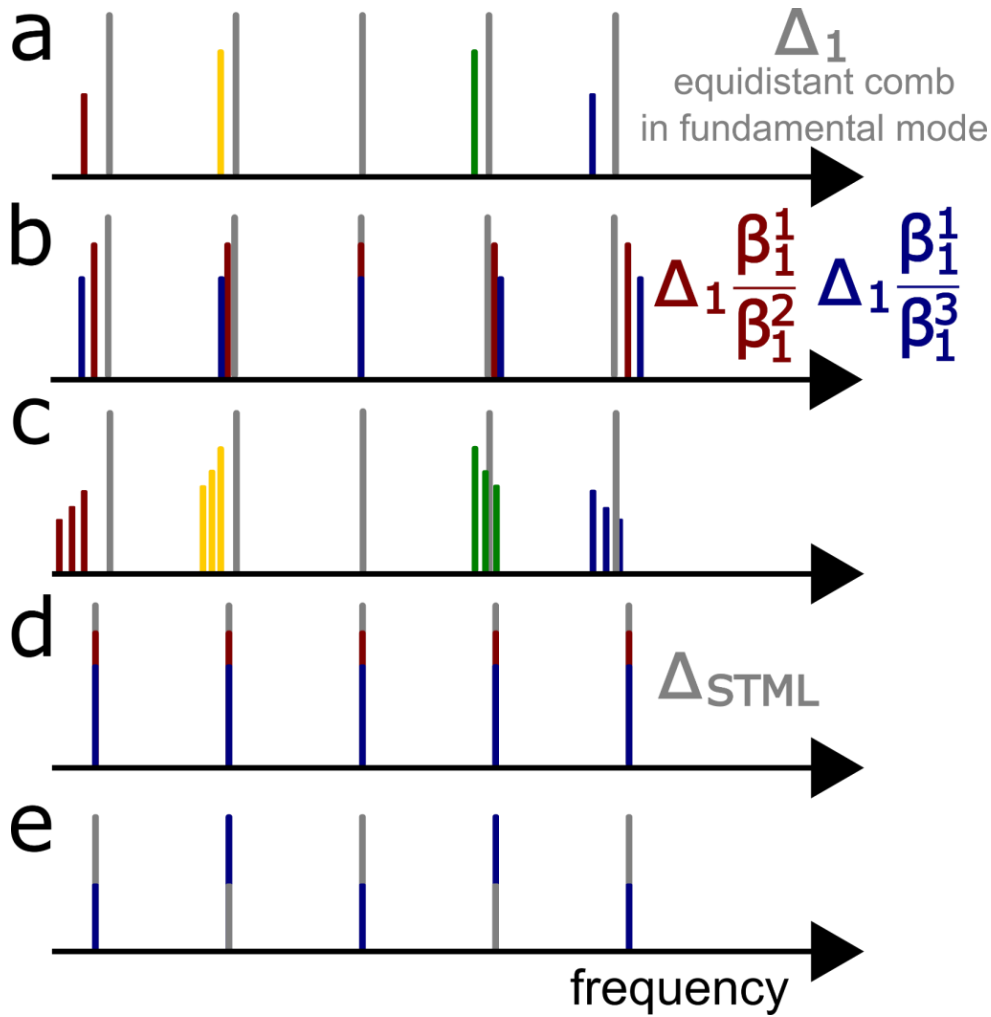

**Supplementary Figure 1: Frequency-domain depiction of spatiotemporal mode-locking. a** shows the effect of chromatic dispersion on the resonances within a given longitudinal mode (the fundamental mode here), **b** shows the effect of modal dispersion, i.e. it shows the resonances within several different transverse mode families, neglecting chromatic dispersion. **c** shows the combined effect of chromatic and modal dispersion, for three transverse mode groups assumed to have an identical resonance at the center frequency, as depicted in b. **d** shows the resonances for a STML first-order eigenpulse, a mode-locked pulse that is identical on each round trip. **e** shows a subharmonic STML mode-locked state, where the modal composition oscillates with a period twice the fundamental cavity round trip. These subharmonic states are closely related to total mode-locking described in early works, and are a special category of multiperiodic STML states described in Section 7 which also include those whose oscillation period is not a sub-multiple of the fundamental cavity period.

**[Supplementary Section 2: Spatiotemporal nonlinear wave equation model]**

Our numerical model is schematically depicted in Supplementary Figure 2.

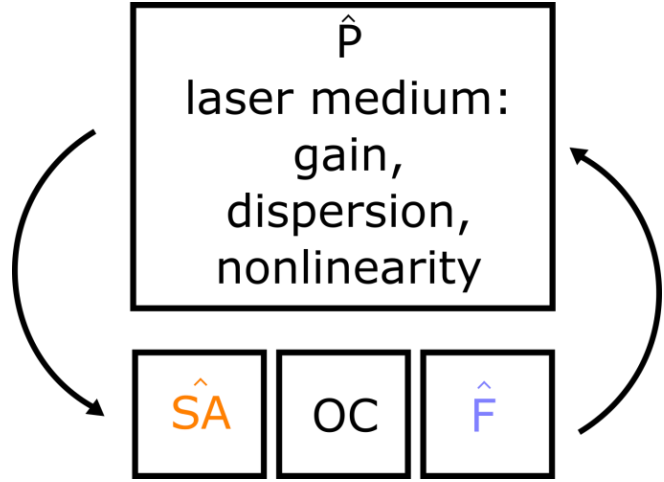

**Supplementary Figure 2: Schematic depiction of the spatiotemporal nonlinear wave equation numerical model of an STML laser.** The laser field propagates in each round trip through the nonlinear, dispersive laser medium, then passes through a spatiotemporal saturable absorber. A portion of the field is coupled out of the laser, and then the field remaining in the cavity is spatially and spectrally filtered before being passed back into the laser medium.

**Propagation through the multimode gain medium in the mode-resolved picture**
In the transverse mode basis, the passive effects of propagation through the multimode fiber medium are described by a generalized multimode nonlinear Schrodinger equation [2–5] for each electric field envelope in each transverse mode, $A_p(t; z)$,

$$\partial_z A_q(t;z) = i\delta\beta_0^{(q)} A_q - \delta\beta_1^{(q)} \frac{\partial A_q}{\partial t} + \sum_{m=2}^{3} i^{m+1} \frac{\beta_m^{(q)}}{m!} \partial_t^m A_q + i\frac{n_2 \omega_o}{c} \sum_{l,m,n} S_{qlmn}^k A_l A_m A_n^* \qquad (S5)$$

where $\delta\beta_0^{(p)}$ and $\delta\beta_1^{(p)}$ are the differences in the propagation constant and group velocity of the transverse mode $q$, $\beta_m^{(q)}$ are its higher-order dispersion coefficients, $n_2$, $\omega_o$, and $c$ are the nonlinear refractive index, the center radial frequency and the speed of light in vacuum respectively. The $S_{qlmn}^k$ are the mode overlap coefficients. For the assumption of only linear polarized modes, it is sufficient to consider the scalar transverse modes $\varphi_q(x,y)$, and the coefficients are given by:

$$S_{qlmn}^k = \frac{\iint [\varphi_q(x,y)\varphi_l(x,y)\varphi_m(x,y)\varphi_n(x,y)]\,dxdy}{\sqrt{\iint \varphi_q(x,y)^2 dxdy \iint \varphi_l(x,y)^2 dxdy \iint \varphi_m(x,y)^2 dxdy \iint \varphi_n(x,y)^2 dxdy}} \qquad (S6)$$

As our experimental oscillator's saturable absorber is realized with nonlinear polarization rotation, this assumption of linear polarization is obviously incorrect. Nonetheless, it constitutes a major simplification, one that has been widely used for many years in modelling single-mode oscillators utilizing nonlinear polarization rotation [6]. In addition to the approximations already mentioned, we are considering a simplified version compared to what is described elsewhere (e.g. Refs. [2–5]), neglecting the effects of self-steepening and stimulated Raman scattering.

This equation is numerically integrated along the $z$ coordinate, with the initial condition at $z = 0$ being the field after the filters (except for the first round trip, where the initial condition is a noise field, roughly approximating the initial noise from which the steady-state field emerges experimentally).

However, the equation described so far does not include a description of the gain within a multimode gain medium. To model multimode gain competition, it is important to incorporate the locality of the saturating gain. We assume that the relaxation time of the gain medium is much longer than the round trip time (for fiber lasers this is an excellent approximation, as the two usually differ by about 4 orders of magnitude), and so the gain responds only to the average power density in the laser. In 3D coordinates, and not considering linear propagation effects, the gain in the multimode gain medium is then approximately described by:

$$\partial_z A(x,y,t;z) = A(x,y,t;z) \frac{g_o(\omega)/2}{1 + \frac{\int |A(x,y,t';z)|^2 \, dt'}{F_{\text{sat}}(x,y)}} \qquad (S7)$$

where $g_o(\omega)$ is the frequency-dependent small-signal gain, and $F_{\text{sat}}(x,y)$ is a saturation fluence related to the distribution of pumped ions in the laser medium. To express this model in the transverse mode basis to be efficiently solved with the GMMNLSE Eqn. S5, we perform a change of basis to Eqn. S7. Multiplying the equation by the denominator of the RHS, we obtain:

$$\partial_z A_p(t;z) + \sum_{l,m,n} S^R_{plmn} \partial_z A_n(t;z) \int \frac{A_l^* A_m}{F_{\text{sat}}} dt' = \frac{g_o(\omega)}{2} A_p(t;z) \qquad (S8)$$

Here, in the approximation discussed earlier regarding linear polarization, the coefficients $S^R_{plmn} = S^K_{plmn}$, and we have assumed a uniform distribution of the gain medium's excited population. As mentioned later, this approximation is valid for the fiber we consider experimentally only when the modes considered are low-order modes, since these modes overlap almost completely with the doped region of the multimode gain fiber we use (non-uniform doping could be incorporated here through a slight modification of the coefficients from $S^R_{plmn}$).

In the integration of Eqn. S5, the contribution to the total $\partial_z A_p(t;z)$ arising from the spatially-saturating gain can thus be found by performing, at each step of the integration along $z$, an implicit calculation of $\partial_z A_p(t;z)$ by solving the linear system of equations

$$T \partial_z \vec{A}(t;z) = \frac{g_o(\omega)}{2} \vec{A}(t;z) \qquad (S9)$$

where $T_{ij} = \delta_{ij} + \sum_{l,m} S^R_{ilmj} \int \frac{A_l^* A_m}{F_{sat}} dt'$ and $\vec{A}(t;z)$ is a vector consisting of the field envelopes $A_p(t;z)$.

In our simulations, we take the fiber length to be 50 cm (shorter, but of similar magnitude to that used in experiments), $g_o(\omega)$ to be a Gaussian function with a 40-nm bandwidth and a peak value of 30 dB, and $F_{sat}$ to be $E_{sat}/A_{eff}$, where $A_{eff} = 1000$ µm$^2$ is the effective area of the fundamental mode of the waveguide considered. If not noted otherwise, we take $E_{sat}$ to be 300 nJ.

**Spatiotemporal saturable absorption**

The spatiotemporal saturable absorber corresponds to some nonlinear optical element that results in a higher loss for regions of low intensity, and lower loss for regions of high intensity. That it is spatiotemporal means it is sensitive to the local spatiotemporal intensity, $I(x,y,t) = |A_i(x,y,t)|^2$. This means that it couples not only longitudinal modes, but also transverse modes, promoting through its differential loss a coherent combination of 3D modes that minimizes the loss through it. Many common saturable absorbers, such as two-dimensional materials or semiconductor saturable-absorbing mirrors, and artificial saturable absorbers, such as the Kerr lens, nonlinear polarization rotation or the nonlinear optical loop mirror, are suitable for acting as spatiotemporal saturable absorbers, provided that where they are placed in the cavity, multiple transverse modes are supported.

In our numerical model, the spatiotemporal saturable absorber is implemented as an ideal transfer function, applied to the field immediately after the end of the gain medium propagation. Unlike other operations, since the full saturation region of the saturable absorber is often relevant, we apply it exclusively in 3D coordinates, with the transfer function

$$A(x,y,t) \to A(x,y,t) \sqrt{1 - \alpha/1 + |A(x,y,t)|^2/I_{sat}} \quad (S10)$$

where $\alpha$ is the modulation depth, and $I_{sat}$ is the saturation intensity. When the other parts of the cavity are described by the mode-resolved picture, the spatiotemporal saturable absorber thus requires performing a change of basis from the modes into Cartesian coordinates, and then backward afterwards.

In our experiments, the spatiotemporal saturable absorber is realized by nonlinear polarization evolution (NPE) within the multimode fiber. While this effect is expected to be very complex, generalizing from the case of single-mode fiber, we anticipate that, for suitable positions of the waveplates within the cavity, nonlinear polarization rotation results in a spatiotemporal loss similar to Eqn. S10. This loss occurs at the output coupler in our experimental laser but depends on the accumulated nonlinear phase shift (and so the amount of nonlinear polarization rotation) through the multimode fiber at each point in spacetime across the field.

In our simulations, $\alpha$ is taken to be 1. We expect that this value is, of course, higher than is achieved in the laboratory using NPE, where values empirically appear to be in the range of 0.5 to 0.95. Our use of $\alpha = 1$ is primarily to speed up the convergence of simulations to the steady-state (which often still correspond to simulating kilometers of propagation through fiber, and thus take days or weeks with efficient parallelized codes [5]). Moreover, qualitatively similar results are found for more realistic values of $\alpha$.

$I_{sat}$ is taken to be 500 kW/$A_{eff}$ = 50 GW/cm², where $A_{eff}$ = 1000 µm² is the effective area of the fundamental mode of the waveguide (corresponding to a 1/e² width of 17.8 µm and a FWHM width of 21 µm). This number is similar to that used in modelling fiber oscillators based on NPE.

**Output coupling**
In our simulations, a fraction of the field is coupled out of the oscillator, with the remaining fraction sent to the spatial and spectral filter. An additional lumped loss, $l$, is included to account for loss within bulk components such as the isolator. Thus, the output coupling step produces

$$A_{out}(x,y,t) = A(x,y,t)\sqrt{OC}$$

$$A(x,y,t) \rightarrow A(x,y,t)\sqrt{l}\sqrt{1-OC}$$

(S11)

In our simulations, we take $l = 0.5$, and the output coupling ratio, $OC = 0.7$.

**Spatial and spectral filtering**
Spatial and spectral filtering are in general spatiospectral filtering, $\hat{F}(x,y,\omega)$, in that the filter may exhibit coupling between its spectral and spatial dissipation. In this work, we focus on the simplified case where $\hat{F}(x,y,\omega) = \hat{F}(x,y)\hat{F}(\omega)$. The spatial filter is applied as a matrix operation,

$$\vec{A}(t) \rightarrow \hat{F}\vec{A}(t) \quad (S12)$$

where the components of the vector $\vec{A}(t)$ are the field envelopes of the field remaining in the cavity after output coupling, and $\hat{F}$ is the spatial filter in the transverse mode basis, consisting of elements $F_{mn} = \langle \varphi_m | \hat{F}(x,y) | \varphi_n \rangle$. The spectral filter is meanwhile mode-indepenent, being applied to each transverse mode envelope in the frequency-domain,

$$\widetilde{A_p}(\omega) \rightarrow \widetilde{A_p}(\omega)F(\omega) \quad (S13)$$

Throughout this work, we use for $F(\omega)$ a Gaussian with a full-width-at-half-maximum of 10 nm, and peak value 1. Spatial filters are meanwhile always Gaussian functions with full-width-at-half-maximum widths and offsets relative to the center of the waveguide noted.

With 3D Cartesian coordinates, spatial and spectral filtering is a straightforward matrix multiplication, in the 3D Cartesian basis instead of the modal basis.

**Propagation through the multimode gain medium in the 3D full-field envelope picture**
For completeness, we also describe the nonlinear wave equation approach in full 3D coordinates. To describe nonlinear pulse propagation in the multimode gain medium in full 3D coordinates, we model the evolution of the 3D field envelope with the equation:

$$\partial_z A(x,y,t;z) = \hat{D}(k_x, k_x, \omega)A + \widehat{W}(x,y,\omega)A + \widehat{N}(x,y,t)A \quad (S14a)$$

where each term is implemented in the domain indicated by its arguments:

$$\widehat{D}(k_x, k_x, \omega) = i\left[\sqrt{\beta_{\text{eff}}(\omega) - k_x^2 - k_y^2} - \beta_{eff}(\omega_0) - (\omega - \omega_0)\frac{\partial \beta_{\text{eff}}}{\partial \omega}\bigg|_{\omega=\omega_0}\right] \quad \text{(S14b)}$$

$$\widehat{W}(x, y, \omega) = i\frac{\beta_{\text{eff}}(\omega)}{2}\left[\left(\frac{n(x,y,\omega)}{n_{\text{eff}}(\omega)}\right)^2 - 1\right] \quad \text{(S14c)}$$

$$\widehat{N}(x, y, t) = \frac{g_0(\omega)/2}{1+\frac{\int |A(x,y,t';z)|^2 dt'}{F_{\text{sat}}(x,y)}} + i\frac{n_2 \omega_0}{c}|A(x,y,t;z)|^2 \quad \text{(S14d)}$$

Here, $\beta_{\text{eff}}$ and $n_{\text{eff}}$ are the propagation constants and of the field at the center of the waveguide.

### [Supplementary Section 3: Operators and attractors of cavity components]

In this section, we describe the form of the projection operations that compose the total cavity round trip operator, $\hat{C}$. For particular effects considered in the paper, we describe the calculation of the attractors of each component's operator, $\hat{O}$, through the solution to Eqn. 2:

$$A(x, y, t) = \lim_{n \to \infty}\left[\hat{R}\hat{O}\right]^n E_0(x, y, t) \quad (2)$$

In Eqn. 2, we consider the (in general nonlinear) projection operation of $\hat{O}$ together with the essential features of the laser cavity, renormalization by the gain, feedback (provided by the recursive repetition of the action through the asymptotic exponent), and the guided oscillating modes of the waveguide (or bounding mirrors, etc.) and their dispersion. The renormalization operation $\hat{R}$ is an idealized saturating laser gain operation, which amounts to the following on an input field $E(x, y, t)$:

1. Calculate the total energy of the field, $U = \iiint |E(x,y,t)|^2 dt$,
2. Multiply the field by $\sqrt{U_o/U}$, so that the energy of the field is restored to a constant $U_o$ regardless of the losses or gains elsewhere.

### Spatial filter

In the main article, we consider the spatial filter attractor without including the spatiotemporal dispersion of the cavity. In this case, it is easy to solve Eqn. 2 by inspection, as described in the main text.

Here, we will consider a slightly more complicated case where $\hat{O} = \hat{F}\hat{T}$. Although this is not the minimal model, it turns out to be insightful in understanding the spatiotemporal effects of the spatial filter (elaborated in Section 5), as well as seeing how the dissected attractors behave. The spatial filter's attractor then represents the best solution to the optimization problem defined by maximizing the transmission through the spatial filter, subject to the constraints of having only the guided transverse mode basis with which to do this, *and* the spatiotemporal coupling due to dispersion in $\hat{T}$. The spatial filter is the simplest of the elements we will consider. As later sections show, to a first approximation the saturable absorber and gain are in direct opposition, and thus the spatial filter play an important role not only in the spatial-filter-driven regime, but also in regimes where the saturable absorber and gain may are most important, since even a weak spatial filter nonetheless serves to bias the eigenpulse's steady-state.

The attractor of the filter with $\hat{O} = \hat{F}\hat{T}$ is still the same as in the simpler case considered in the main article, with $\hat{O} = \hat{F}$. Taking $F$ to be the filter in the transverse mode basis of the cavity, with elements $F_{mn} = \langle \varphi_m | F(x,y) | \varphi_n \rangle$, where $F(x,y)$ is the transmission function of the filter, and $\varphi_i = \varphi_i(x,y)$ are the transverse modes. The matrix $F$ has eigenvectors $\vec{v}_i$, which are complex mode coefficients for fields that pass through the filter with only a uniform loss, i.e. $\hat{F}\vec{v}_i = t_i \vec{v}_i$, where $T_i = |t_i|^2$ is the energy transmission through the filter. Since $F_{mn}$ depend only on $x$ and $y$, the solution is likewise independent of time. If the initial field $E_0(x,y,t)$ is a uniform noise field, then the solution to Eqn. 2 for the spatial filter is its global attractor, the highest-transmission eigenmode of the spatial filter (the eigenmode for which $T_i = |t_i|^2$ is the largest). Thus, the attractor of the spatial filter is a field composed of, at all points in time $t$, the complex amplitudes of the highest-transmission eigenmode of the spatial filter matrix, in the basis of the cavity's guided modes. Due to spatiotemporal coupling caused by $\hat{T}$, the global attractor of Eqn. 2 for the spatial filter is a continuous-wave field, $\partial_t A(x,y,t) = 0$ for all $x, y$.

The behavior of Eqn. 2 for the spatial filter contains several simple, but potentially subtle aspects, that illustrate features of the laser cavity's selection of a steady-state solution. Before moving on to other effects, we will highlight these here. Although Eqn. 2 for the spatial filter is easily solved by inspection, it is instructive to consider the convergence to the field to the attractor with a numerical solution of the equation.

Supplementary Figure 3 shows the result of performing a numerical calculation of the solution to Eqn. 2, for a Gaussian spatial filter of full-width-at-half-maximum shown above each plot, with $\hat{T}$ and the mode basis the 7 modes and fiber considered throughout the rest of this work (and in Ref. [7]). The four plots illustrate how the "attractiveness" of the spatial filter's attractor (shown in each plot) depends on the transmission gradient and the modal dispersion in the fiber (which affects how much a non-CW field will distort from the attractor due to the action of $\hat{T}$). At any given point in time, the transmission gradient can be considered through the difference between the highest-transmission and next-highest-transmission spatial filter eigenvalues, $T_1$ and $T_2$, normalized to the highest-transmission, i.e., the gradient proxy is $(T_1 - T_2)/T_1$. For the 6 μm filter, this value is 0.97, while for the 40 μm filter it is 0.16. As this value increases, the extent to which the highest-transmission eigenmode of the filter (the attractor of the spatial filter in our nomenclature) is promoted over other eigenmodes increases. As the modal dispersion in the cavity increases, the effect of the cavity linear propagation operator $\hat{T}$ increasingly distorts pulses which are not temporally uniform. This results in a higher loss in the spatial filter, and thus a more rapid selection of the attractor, a temporally-uniform field across the temporal window of the simulation.

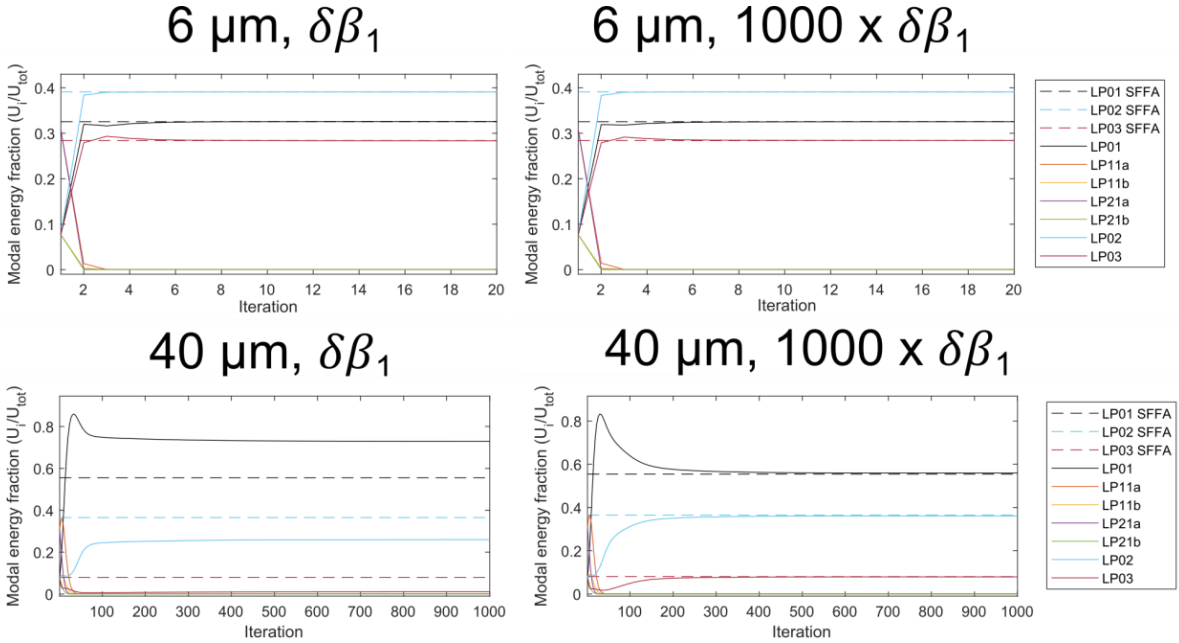

**Supplementary Figure 3: Numerical simulations of Eqn. 2 for the spatial filter.** The plots show the modal energy fraction (note the attractor of the spatial filter is an attractor for the electric field; this representation in terms of energy suppresses the amplitude of the modal coefficients) with each round trip, or iteration of Eqn. 2. The plots show that, for both small and large modal dispersion, the field rapidly converges to the attractor for a small, 6 µm FWHM Gaussian filter, for which the gradient proxy is 0.97. For a 40 µm FWHM Gaussian filter, the attractor is weaker as the gradient proxy is 0.16. This is evident by the much slower attraction for both modal dispersion cases (note the x-axis in the bottom panel). However, in the case of large modal dispersion, the field nonetheless converges much more rapidly than for small modal dispersion. For the sake of extending the number of iterations required for convergence enough to be visible in the upper row, the field has been initialized as a noise field with a transverse mode composition far from the attractor. SFFA = spatial filter's field attractor.

To continue our use of the spatial filter as a simple concrete example of our approach, several other remarks are worthwhile. These highlight nuances that are similar to those exhibited by the other effects considered subsequently.

First, despite not depending on time, the spatial filter's attractor nonetheless exerts an influence on pulse-shaping in the time domain (see Supplementary Section 5). In absence of spatiotemporal coupling in the rest of the cavity, the spatial filter's attractor would be any field that, at all points in time $t$, has the modal coefficients of the filter's lowest-loss eigenmode. However, modal dispersion causes the different transverse field components to disperse each round trip, and thus a pulse, or any temporally non-uniform field initially fully matching the attractor will diverge from it. The action of the spatial filter, in the presence of spatiotemporal coupling, is to spatially homogenize the field across time, since at each point in time it promotes a single modal composition. In mode-locked operation, this homogenization corresponds to a shortening of the pulse duration, which acts as a (dissipative) compensation for modal walk-off.

Second, like all the attractors we consider, it is limited by the representational capacity of the cavity's guided transverse modes. The spatial filter's attractor is the highest-transmission eigenfunction of the filter decomposed into the transverse mode basis of the cavity. This is the configuration of the field with maximum transmission through the filter allowed within the representation.

Third, possible solutions to Eqn. 2 are limited to pure eigenfunctions of the filter (excepting possible higher-order solutions where the field oscillates each round trip). The laser optimization problem is one in which the self-consistency is an essential constraint; the minimal dissected attractor models capture this. In this case, a field that consists of a superposition of the filter's eigenfunctions will, on each application of the filter, have its components in each eigenfunction modified by the transmission eigenvalues of the eigenfunctions. All eigenfunctions except the highest-transmission one are metastable solutions, as presuming the amplitude of the field in any higher-transmission eigenfunctions is zero, the highest-transmission *occupied* eigenfunction will dominate all others. This discrete quality to the solutions of Eqn. 2 is one example that emphasizes how the single-optimum depiction used in the EEE surface cartoons used throughout the main article are significantly simplified (in addition to being plotted in one-dimension rather than the 2M degrees of freedom available to the field). This discreteness of solutions emphasizes how the apparent continuously-smooth shape of the EEE surface does not necessarily lead to continuous changes in the eigenpulse parameters when cavity parameters are changed. Typically, instead we observe small changes due to solution-pulling (i.e., slight perturbation of the maximum of EEE) from a strong attractor, until a critical value of the change, where a rapid transition to a qualitatively-distinct eigenpulse occurs. This behavior is as expected for the EEE surface being the sum of primarily convex EEE curves from the distinct components, i.e. from the composition of a small number of strong attractors.

Last, the spatial filter's eigenfunctions consist of amplitudes for the electric field modal components, meaning the relative phase between transverse modes is equally important as their relative amplitude. This means that the spatial filter's attractor is a phased, coherent combination of multiple transverse modes, and so the spatial filter helps to promote coherence in the field, whether it is multimode or not.

**Spatiotemporal saturable absorber**
The spatiotemporal saturable absorber's attractor represents the solution to the optimization problem defined as maximizing the electric field intensity (at all points in time) inside the spatiotemporal saturable absorber, subject to the constraints of the cavity's linear dispersion/transmission effects, the guided transverse mode basis, and the available gain and gain bandwidth. To a first approximation, this corresponds to forming a pulse, minimizing its duration and establishing a transverse mode composition that minimizes (on average across the pulse in time) the effective area,

$$A_{\text{eff}} = \frac{(\iint I(x,y)\,dxdy)^2}{\iint I(x,y)^2\,dxdy} \qquad (S15)$$

The spatiotemporal saturable absorber's attractor is found by taking $\hat{O}$ in Eqn. 2 to be the operation in Eqn. S10. To solve Eqn. 2, we choose the saturable absorber parameters to be the same as those used in our full simulations and vary the parameter $U_o$. We initialize the field

$E_0(x, y, t)$ with noise equally distributed among the modes and iterate until a converged solution is found. For the saturable absorber function we use, we find that this calculation often leads to an oscillatory steady-state. While this is a notable feature of the saturable absorber's attractors, for present purposes we are interested in static solutions. To understand the role of the saturable absorber in STML eigenpulses that are not oscillating beyond the fundamental repetition rate (i.e., subharmonic or multiperiodic), we will here only consider on the solutions that converge to a pulse whose period equals the cavity period (i.e., fixed points or first-order eigenpulses).

Supplementary Figure 4 shows the transverse mode composition of the calculated spatiotemporal saturable absorber field attractor (SAFA) versus the ratio of the peak intensity, $\max[I(x, y, t)]$ of the attractor, to the saturable absorber's saturation intensity. It is evident that there are roughly two classes of SAFA, those in the unsaturated or 'low intensity' regime, and those in the saturated or 'high intensity' regime.

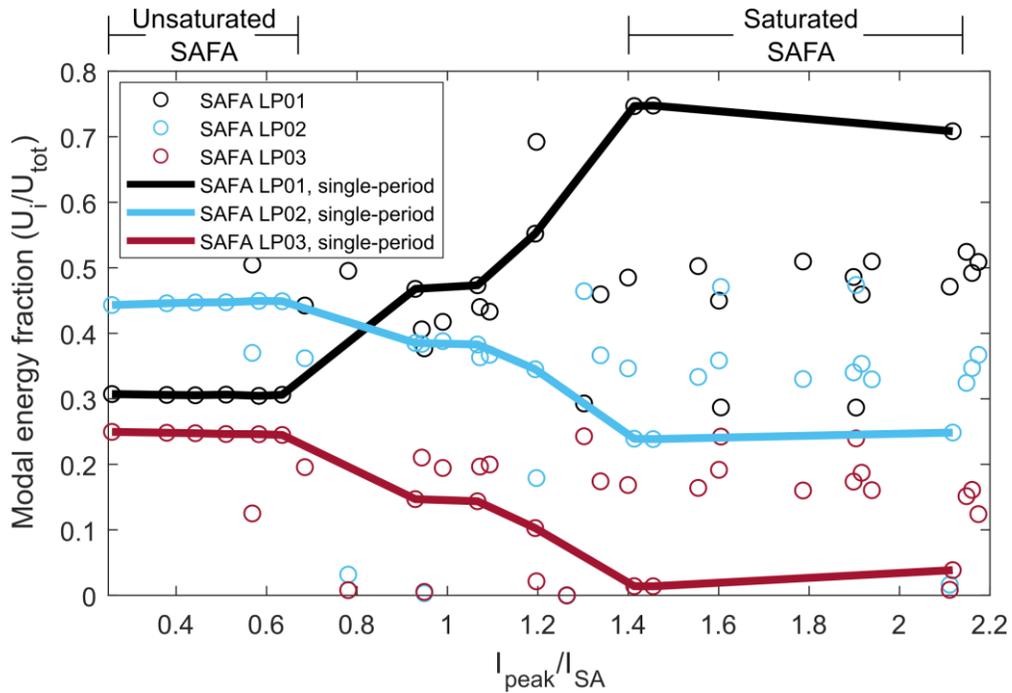

**Supplementary Figure 4: Spatiotemporal saturable absorber field attractor**. The plot shows the transverse mode composition of the solution to Eqn. 2 with $\hat{O}$ the saturable absorber transfer function defined by Eqn. S10, and the linear spatiotemporal dispersion of the cavity, $\hat{O} = \widehat{SA}\hat{T}$. Solutions are found by varying the renormalization energy but are plotted here as a function of the peak intensity of the field relative to the saturable absorber's saturation intensity, since this produces the most informative trend. The solid lines show the trend for solutions characterized by static fields, while other points correspond to oscillating solutions. The solid lines show that there are roughly two main regimes, the unsaturated attractor, and the saturated attractor, depending on whether the pulse's peak intensity is smaller or larger than the saturation intensity of the saturable absorber.

**Gain**

The gain's attractor is the field that maximizes the extraction of the energy in the gain medium, by maximizing the energy density of the field in its 3D volume. If $F_{sat}(x,y) = F_{sat}$, then the gain's attractor is essentially the maximizing of $A_{eff}$ at each point in time.

Due to the integration over the temporal window associated with the slow gain dynamics, the gain's attractor does not depend on time except through the spectral bandwidth in the function $g_o(\omega)$. In contrast to the spatial filter, the gain's effect on the time domain occurs through this *spectral* filtering, even in the presence of the spatiotemporal coupling through $\hat{T}$. Its slow response means it is only sensitive to the average mode composition across the entire time window. However, its spectral filtering causes a (relatively weak) attraction towards a CW field. Thus, its attractor is a continuous-wave field, spectrally centered at the peak of $g_o(\omega)$, with a transverse mode composition that maximizes the overlap of the field with the available energy, described by $F_{sat}(x,y)$.

To describe the gain's attractor, we may take $\hat{O}$ in Eqn. 2 to be the integration of Eqn. S8 over the fiber length from $z = 0$ to $z = L$, i.e., $\hat{O} = \hat{P}(\hat{T}, \hat{G})$ Since the gain is nonlinear and occurs simultaneously with the linear propagation effects within the fiber, it is not necessarily a good approximation to treat the multiplication by $\hat{T}$ as a factorable operation. Instead, we integrate the equations described in Section 2, except neglect Kerr nonlinearity.

Supplementary Figure 5 shows the convergence of initial noise fields to the gain's attractor. A parameter that determines the rate at which the field converges to the gain's attractor is the ratio of the rescaling energy to the gain saturation energy (in a realistic oscillator model this would be roughly the steady-state gain per round trip). When this ratio is small, the field experiences little gain each round trip, and so the field converges towards a CW field slowly. However, gain competition effects manifest strongly, since the gain for spatial fields different from the attractor are strongly suppressed through cross-gain modulation that occurs when the gain is saturated. Consequently, the transverse component of the attractor is realized quickly. For a large ratio, significant gain occurs each round trip, and convergence to a CW field is fast. However, saturation effects are minimal so the transverse component of the attractor manifests more gradually. If we translate this observation to a real cavity, it implies that the importance of the gain's attractor in the full cavity depends on the loss of the field each round trip (e.g., the output coupling ratio), and of course, on the gain saturation energy.

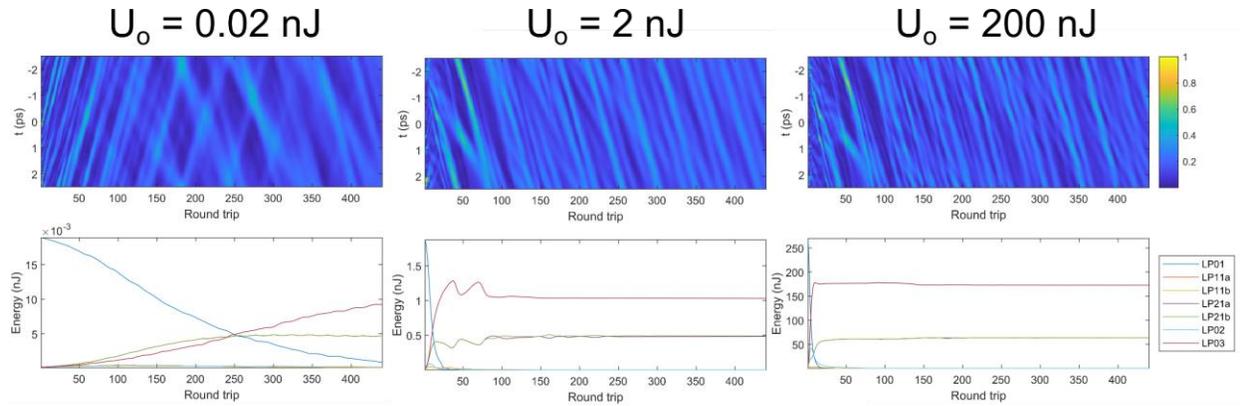

**Supplementary Figure 5: Convergence of the field to the attractor of the gain.** For different values of the rescaling energy, the convergence rate is different, but the field converges to roughly the same, area-maximizing, continuous-wave field. For all three cases, the gain saturation fluence is 300 nJ / 1000 µm². In all three instances, we have not taken the simulations all the way to a pure continuous wave field. However, the upper plots show that a temporally uniform field is the asymptotic limit.

### Saturable absorber and gain (SAGE)

We can also let $\hat{O}$ contain multiple effects, in order to describe attractors that depend on interactions or competitions among multiple distinct effects in the cavity. This is explored in greater detail in the next section, but here will apply it to the 7-mode simplified cavity to obtain the field attractors for the SAGE regime. Here, we take $\hat{O} = \widehat{SA}\hat{P}(\hat{T}, \hat{G})$. The SAGE attractor depends strongly on the amount of gain saturation and the intensity of the pulse formed relative to the saturable absorber's saturation intensity. As a result, when we vary the rescaling energy $U_o$ relative to the 300 nJ / 1000 µm² saturation fluence of the gain, we find significant differences in the steady-state field distribution among the modes. In addition, we find that the spatial filter influences the details of the SAGE attractor, especially in in the strongly saturated and strongly unsaturated regimes

In our full simulations of the 7-mode cavity in the SAGE regime (see Section 5), we find that the typical energy of the pulse entering the gain fiber is about 2-6 nJ, corresponding to the boxed region in Supplementary Figure 6. It is from this region that we obtain our estimate of the modal composition of the SAGE attractor for Figure 2 of the main article (specifically, we average across the four values in the dotted box in Supplementary Figure 6 to obtain the modal composition lines plotted in Figure 2).

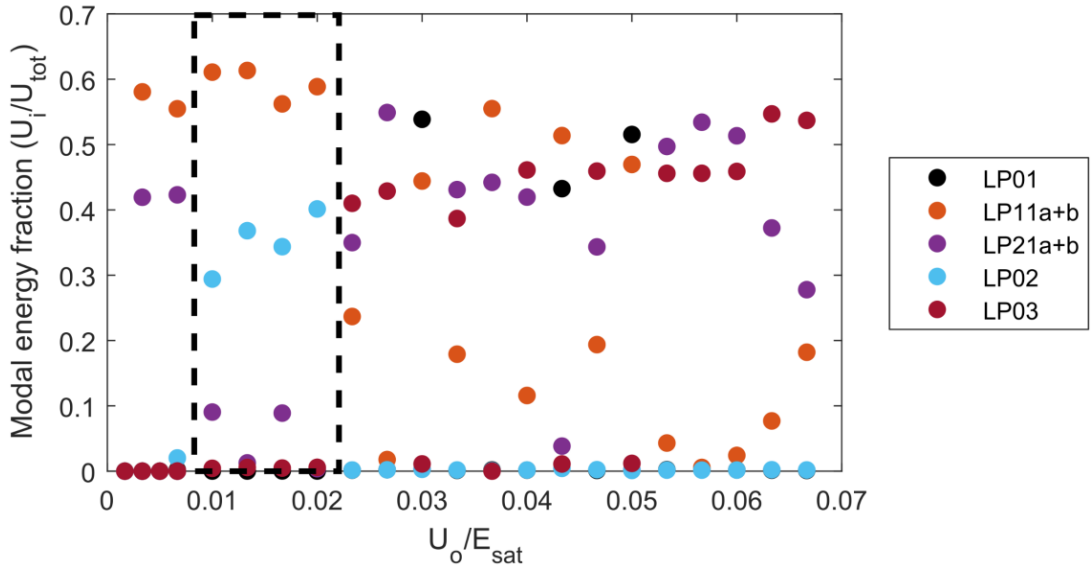

**Supplementary Figure 6: SAGE attractor in the 7-mode cavity.** The points show the modal energy fraction (energy in the indicated mode family divided by the total) for varying the rescaling energy to gain saturation energy. The saturable absorber parameters are fixed to the values considered in the previous section: modulation depth of 100%, and saturation intensity 500 kW/$A_{eff}$ = 50 GW/cm$^2$, where $A_{eff}$ = 1000 μm$^2$.

### Nonlinear and dispersive propagation, and the spectral filter

In this work, we do not consider the independent attractors of these effects explicitly. As is evident in Supplementary Section 5, this neglect is justified in that these effects collectively counterbalance one another in normal dispersion mode-locking. In each round trip, the pulse broadens in time and spectrum, acquiring an approximately linear frequency chirp. The spectral filter undoes the spectral broadening from nonlinearity, and, due to the frequency chirp, reduces the pulse duration by filtering off the temporal wings of the pulse. This closed balance between these effects means that they cancel each other out with respect to influencing the steady-state pulse parameters. For this reason, and because the other effects are more important in phenomenon that are novel to STML, we focus our attention on the other intracavity effects.

This neglect is however, an approximation. All three effects have roles in three-dimensional mode-locking that are not similar to one-dimensional mode-locking. For example, nonlinear cross phase modulation between transverse modal components causes changes in the spectral center-of-mass in each modal component of a pulse, resulting in a change in the walk-off rate of those components because of group velocity dispersion. Depending on the sign of the modal dispersion, this results in either an increase in the walk-off between a pulses' modal components, or a reduction. In the former case, the spectral filter then effectively compensates for modal walk-off, while in the latter it does not.

Furthermore, this approximation is likely to be very poor for other for certain interesting cases of STML, and our consideration in this work has centered on regimes of STML where the approximate cancellation of these effects by one another permits a focus on other effects. Many useful and novel regimes of STML will likely make explicit use of these effects as primary driving forces for the steady-state pulse. For example, we expect that just as in one-dimensional, single-mode gain media, the nonlinear propagation in three-dimensional multimode gain media should have exact or approximate attractors that correspond to three-dimensional or multimodal nonlinear solitary waves. In space-time, multimode solitons occur for anomalous dispersion, while in space an attractor towards the fundamental mode is expected [8–12]. Additionally, related nonlinear waves like similaritons might also act as spatiotemporal attractors of nonlinear propagation in dissipative multimode media.

**[Supplementary Section 4: Reduced laser models]**
While our primary numerical model in Section 2 combines almost all relevant effects together, including nonlinearity, dispersion, and gain simultaneously in propagation through the medium, we can extend the approach of the last section to build a hierarchy of more advanced, but still approximate, efficient and insightful, laser models. These models can be advantageous for several reasons, including that they are numerically much more efficient when many modes are relevant. These models are also easier to interpret than more complete, but more complex, models like those in Section 2. Last, they allow for effects like disorder to be introduced somewhat more easily than in the models described in Section 2. In sum, they help to fill in the gaps in understanding between our experiments and few-mode nonlinear wave equation simulations, reconciling several complicating effects present in the experiment with the simplified dissected attractor laser models described earlier and emphasized in the main text.

As an example, to model the experiments in Fig. 3 in the main article, we neglect the details of the nonlinear pulse propagation, but consider all other effects. For this, we solve

$$A(x,y,t) = \lim_{n\to\infty} [\hat{R}\hat{T}_r\hat{F}\widehat{SA}\hat{G}]^n E_0(x,y,t) \qquad (S16)$$

where $\hat{T}_r$ is the linear transmission through the cavity, including disorder. $\hat{R}$ is obviously artificial here and is not strictly required in reduced laser models; it is applied here practically to minimize the need to systematically vary the gain saturation energy when other parameters are modified. Here, we use a simplified model of the gain operator to speed up calculations. We take $\hat{G}$ to be the approximate solution to the full-field gain saturation differential equation S7:

$$\tilde{E}(x,y,z+L,\omega) \to \tilde{E}(x,y,z,\omega) \exp\left[\frac{Lg_o(\omega)/2}{1+\frac{\int |E(x,y,z,t')|^2\,dt'}{F_{\text{sat}}(x,y)}}\right] \qquad (S17)$$

The expected distribution of $F_{sat}(x,y)$ is a constant for $\sqrt{x^2+y^2} \leq 35$ µm, and zero everywhere else. When the gain is applied as per Eqn. S17, this harsh function leads to sharp edges that are not observed if equation S7 is integrated across the fiber. To compensate for this, we take $F_{sat}(x,y)$ to be a constant, and multiply $g_o(\omega)$ by a spatial distribution correction function, $G_S(x,y) = \exp\left[-\left(\sqrt{x^2+y^2}\right)^4/w_{\text{dopant}}^4\right]$.

Disorder is implemented in $\widehat{T}_r$ by multiplying the vector of modal coefficients at each frequency before and after applying the spatiotemporal dispersion by a matrix $M_D = I + rX$, where $I$ is the $M$ by $M$ identity matrix, $r$ is a parameter characterizing the strength of disorder (between 0 and 1) and $X$ is a random complex matrix (drawn from a uniform distribution from -1 to 1 for real and imaginary components). As experimental disorder manifests both in conservative (coupling between guided modes) and dissipative coupling (coupling to radiating modes), $X$ is not symmetric. We implicitly assume disorder only couples transverse modes, and take the disorder to be fixed across the timescale of the simulation. Thus, we apply two different, fixed matrices $M_D^{(1)}$ and $M_D^{(2)}$ before and after applying the spatiotemporal dispersion each round trip. For the simulations shown in Fig. 3, we additionally take the output field to be the rejected light from the saturable absorber, as in the experiment.

We find that disorder is necessary to adequately describe experiments. Supplementary Figure 7 illustrates this, showing how the pulses formed in the broad-filter regime only come to resemble the experimental pulses (see Supplementary Figures 16 and 18) once a significant amount of disorder is included. We note that experimentally such a large value of $r$ is surprising. However, we attribute significant linear mode coupling to the clamps used to hold the fiber firmly fixed to alignment stages. We find that even when clamps are maintained at the minimum tightness required to hold the fiber in place, they still introduce noticeable coupling among many modes. In the future, we anticipate that use of large end caps on the fiber will eliminate this source of distorting linear mode coupling. However, as-is, the strong mode coupling induced by the clamps illustrates the surprising adaptability and robustness of STML.

Although we find for reasonable parameters, we obtain qualitatively similar results as experiments, we found that somewhat different parameters led to the best match with experiments in each regime. This is not unexpected, given that, e.g., we expect stronger disorder among high-order modes (in contrast to the uniform assumption of the model), we experimentally need to adjust the pump power and saturable absorber conditions to achieve mode-locking in each regime, resulting in slightly different parameters in each case. For the reduced model simulations in Fig. 3, for the narrow spatial filter simulation we used a 6 µm FWHM Gaussian spatial filter, offset from the center by 1 µm, a 40 µm gain dopant width, $r = 0.05$, and a saturable absorber saturation intensity of 40 GW/cm², and gain saturation fluence of 3 mJ/cm². For the intermediate filter simulation, these parameters were respectively 50 µm, 2.5 µm, 35 µm, 0.15, 30 GW/cm², and 3 mJ/cm². Finally, for the broad filter simulation, the parameters were were 1000 µm, offset by 270 µm, 35 µm, 0.2, 25 GW/cm², and 10 mJ/cm². We plot a series of additional visualizations of the eigenpulses predicted by the reduced model in these regimes in Supplementary Figure 8.

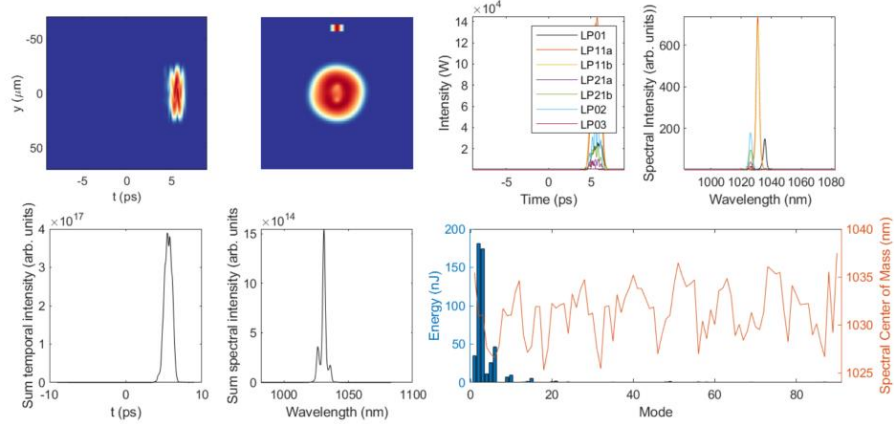

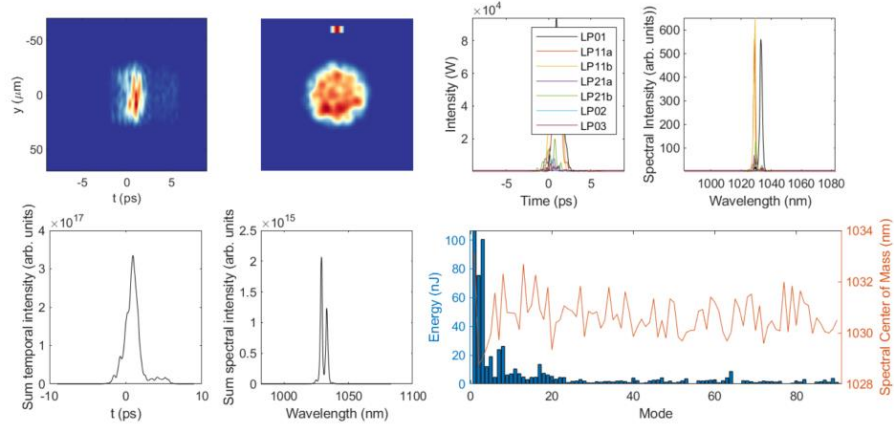

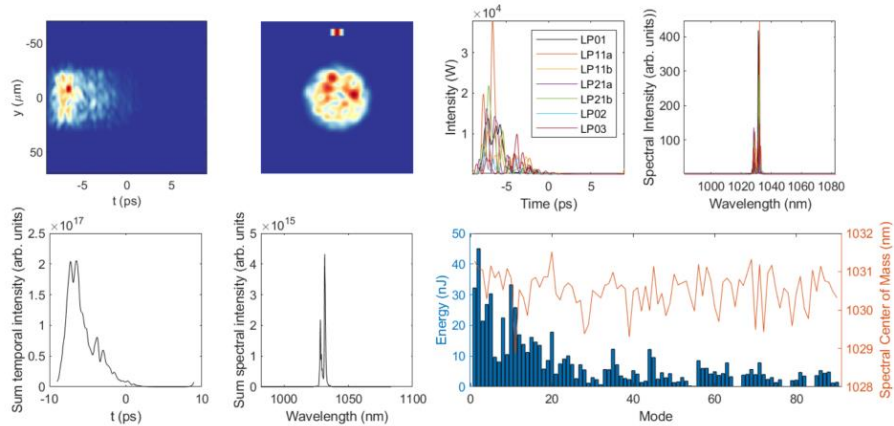

**Supplementary Figure 7: Example calculation from a reduced model for a broad spatial filter, with varying random linear mode coupling and all 90 transverse modes.** In each subpart, the plots show (from top left clockwise to bottom right) the x-integrated spatiotemporal intensity, the beam profile, the mode-resolved temporal intensity (only the 7 modes considered in most of our GMMNLSE sims are shown for clarity), the mode-resolved spectral intensity, the space-integrated temporal intensity, the space-integrated spectrum, and the distribution of energy and spectral center of mass among the modes.

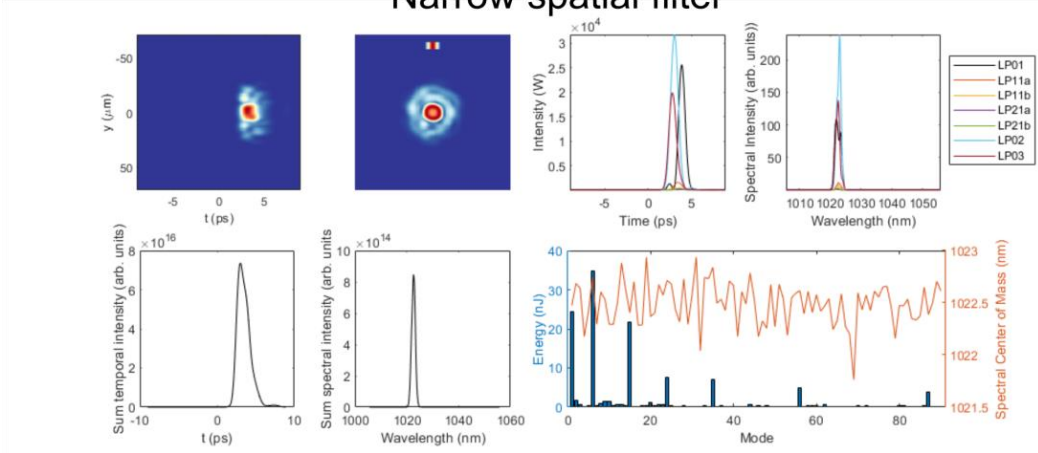
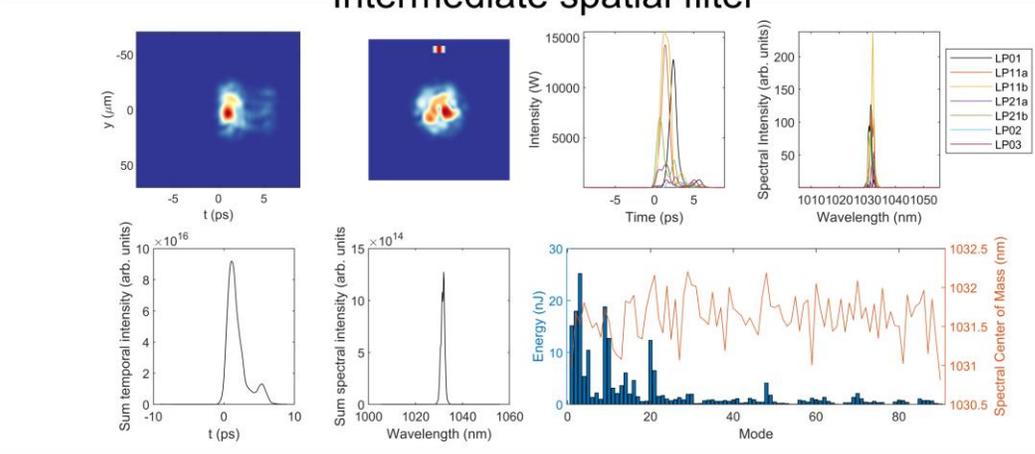
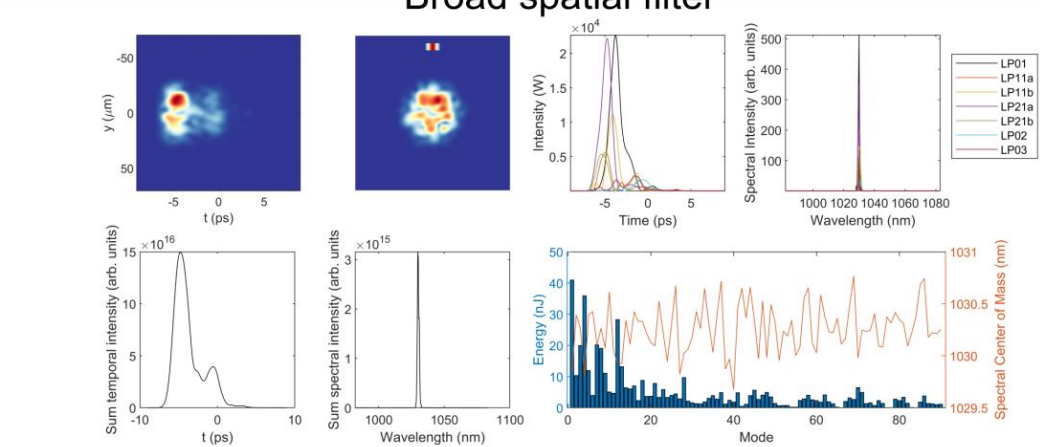

**Supplementary Figure 8: Example calculation from a reduced model including linear random mode coupling and all 90 transverse modes, for three different spatial filter sizes.** In each subpart, the plots show (from top left clockwise to bottom right) the x-integrated spatiotemporal intensity, the beam profile, the mode-resolved temporal intensity (only the 7 modes considered in most of our GMMNLSE sims are shown for clarity), the mode-resolved spectral intensity, the space-integrated temporal intensity, the space-integrated spectrum, and the distribution of energy and spectral center of mass among the modes.

**Analysis of experimental fiber parameters with respect to SAGE attractor**

To understand these results (as well as the experimental results they resemble), we examine the gain fiber's properties to estimate the characteristics of the attractors of the SA and gain in the full experimental fiber, and so the eigenpulses in the SAGE regime. Supplementary Figure 9 shows the calculated effective area, the effective area overlapping the doped region, and the modal dispersion parameter, for all 90 transverse modes of the fiber used in our experiments. These calculations are based on our best-estimate of the fiber profile. The areas illustrate that, while the lowest-order modes (up to about 25) exhibit a high overlap with the doped region, many much higher-order modes, up to mode 90, exhibit a roughly similar overlap. We thus expect the gain's attractor for such a fiber to contain a large number of modes, predominantly lower order modes up to about mode 25, but also non-negligible contributions from many other even higher-order modes.

The modal dispersion shows that, while the lowest-order modes do comprise a set of relatively low-modal-dispersion modes, their average velocity is not especially different from many much higher order modes. In the SAGE regime, the saturable absorber's influence is to promote the eigenpulse composition with a minimal temporal broadening rate, and thus to comprise modes with as little relative modal walk-off as possible. While in a few-mode fiber without disorder, a degenerate mode family is the natural choice, in a fiber with many modes and non-negligible linear mode coupling between them, the saturable absorber instead chooses a pulse in a mode configuration that might be thought of as resembling a cavity principal mode, or perhaps a cavity "super principal mode" [13–16].

Integrating these considerations together, in the 90-mode experimental fiber, the SAGE pulses manifest as large-area, highly-multimoded pulses that, while dominated by the low-order modes, also contains a large amount of energy distributed among the highest-order modes. Since the modal dispersion experienced by the pulse is minimal, the effects of cross-phase and cross-amplitude do not produce asymmetric spectral broadening/filtering, as would occur if the pulse was comprised of modal components moving away from each other. This is evident by the relatively small variation of the spectral center-of-mass across the modes.

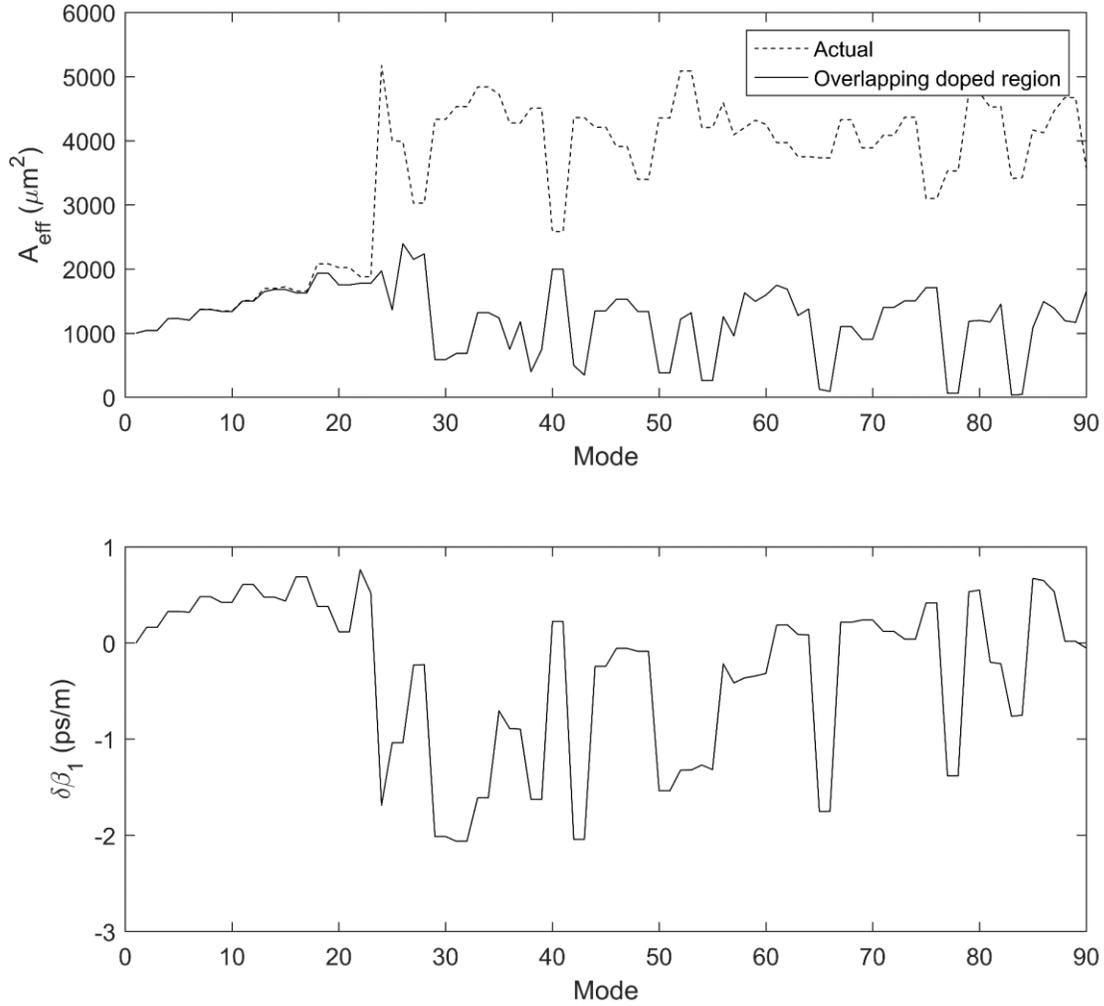

**Supplementary Figure 9: Calculated characteristics of modes in the fiber used in our study.** The top plot shows the effective area of the modes, as well as the effective area calculated only in the doped region of the fiber. The bottom plot shows the distribution of modal dispersion, the walk-off of each transverse mode relative to the fundamental. Both parameters suggest that, in the SAGE regime, a many-moded MM field should be expected, including predominantly energy in the lowest 25 modes, but also non-negligible energy in higher-order modes.

**[Supplementary Section 5: Characteristics and pulse evolutions in the different regimes of STML for varying spatial filter size]**

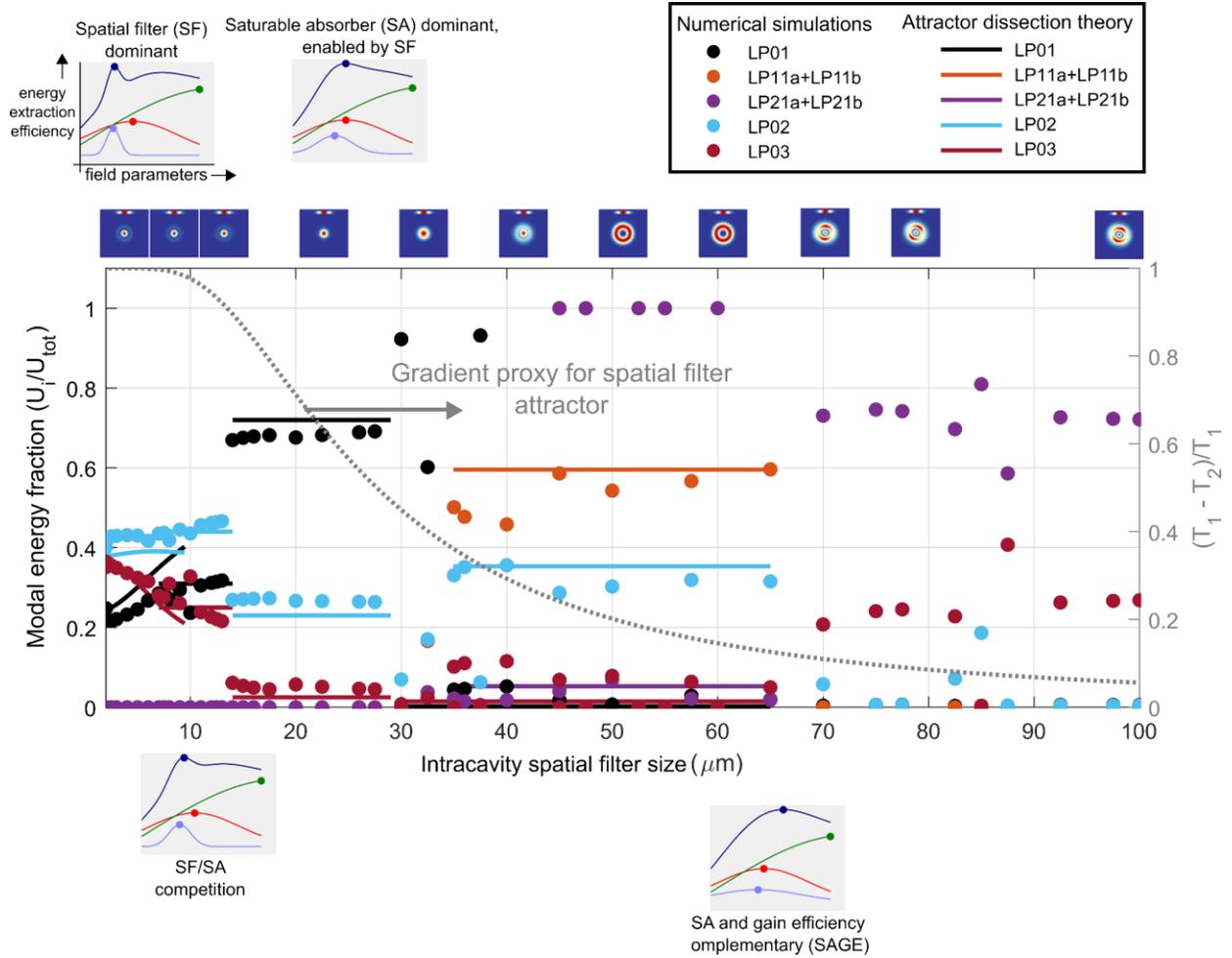

**Supplementary Figure 10: Extended version of Fig. 2.** The figure shows the same data as is shown in Fig. 2 of the main article, except the x-axis is extended to show the largest filter-size region considered, where multimode oscillating SAGE pulses are the most common steady-state.

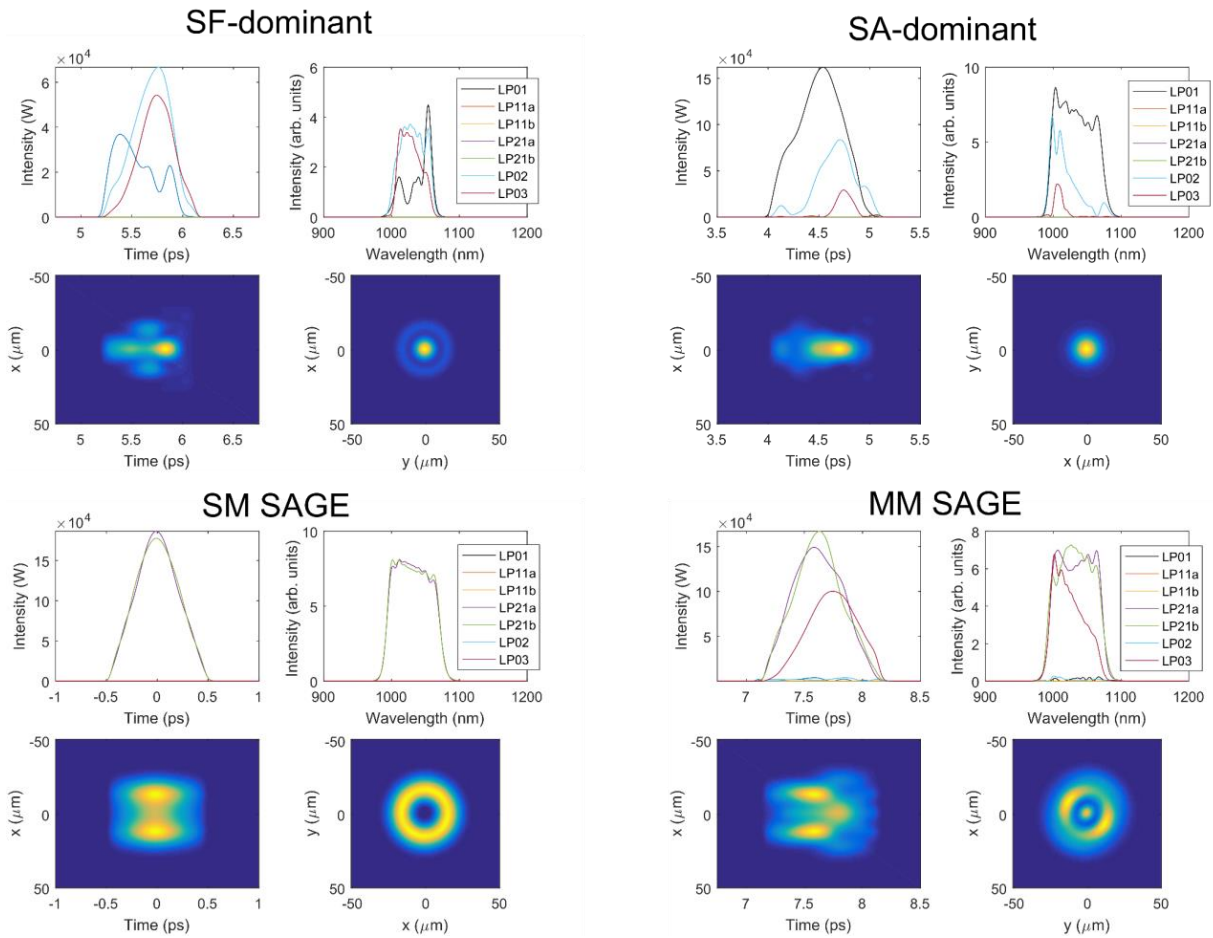

**Supplementary Figure 11: Steady-state pulses in 4 distinct operating regimes of the STML oscillator considered in the few-mode numerical model outlined in Section 2. Each quadrant** shows, for each regime, the mode-resolved spectrum and pulse intensity at steady-state, the corresponding beam profile (integrated over time) and the space-time intensity (integrated along one spatial dimension).

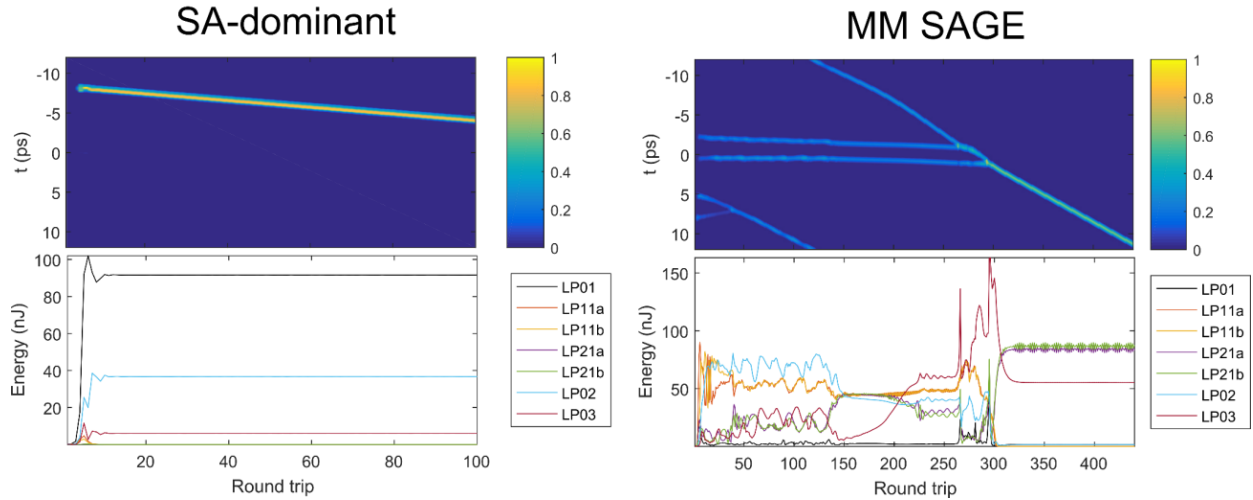

**Supplementary Figure 12: Examples of route to steady state from noise.** The two sets of plots illustrate examples of the route to steady-state from an initial noise field for similar cavities with different spatial filter sizes. The top part of each plot shows the temporal intensity (integrated over both spatial dimensions), while the bottom shows the energy in each transverse mode. For the MM SAGE, several collisions occur prior to a single pulse remaining, similar to what occurs during the formation of single-mode SAGE pulses. For MM SAGE pulses, oscillations occur in the steady-state solution, which can be observed in the energy plot. See also the second part of the Supplementary Movie.

## Steady-state evolution in the spatial filter dominant regime

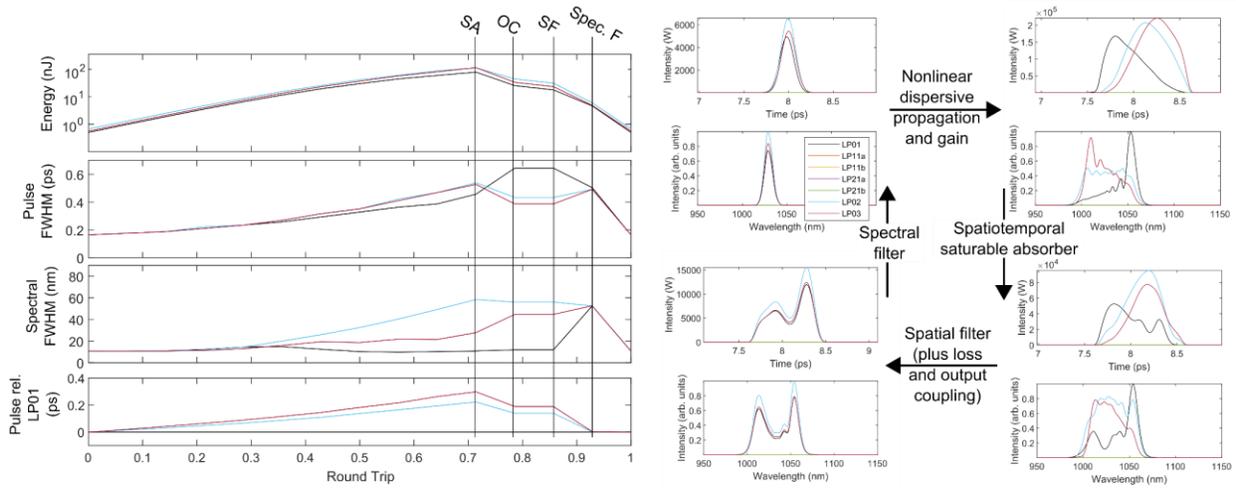

## Other steady-state intracavity pulse evolutions

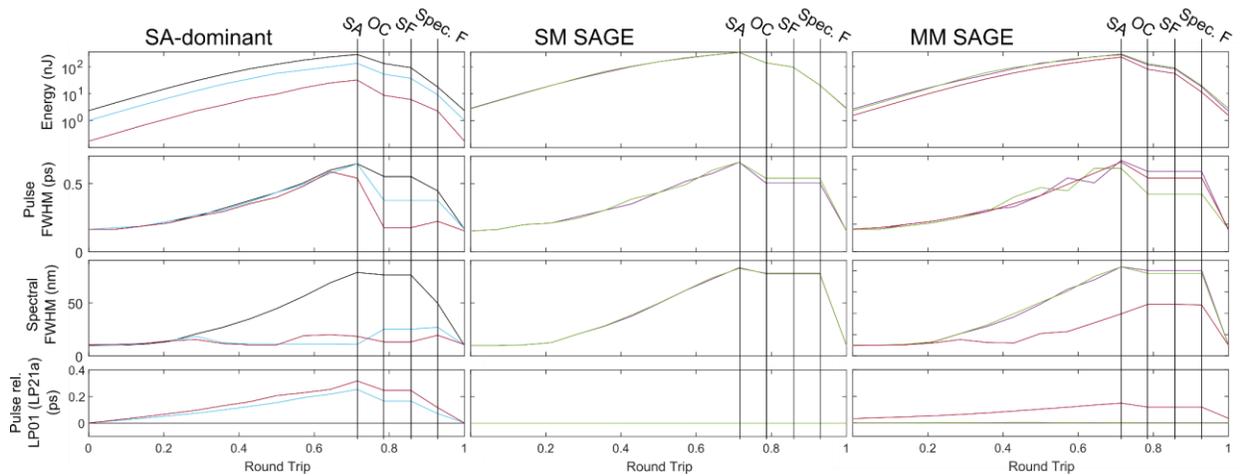

**Supplementary Figure 13: Intracavity spatiotemporal pulse evolutions in the 4 regimes. The top section** shows the steady-state pulse evolution in the spatial-filter-driven regime. The left plot shows the evolution of different parameters of the field across the round trip. The round trip is broken up into 14 'pages', with the first 10 corresponding to equally-spaced segments within the gain medium, and the last 4 the field after the indicated effect has taken place. The right plots in this section show the temporal intensity and spectral intensity in each transverse mode at relevant points in the evolution. **The bottom section** shows the steady-state pulse evolutions for the other four regimes considered. While the MM SAGE regime does exhibit oscillations beyond a single round trip, the details of the evolution do not change significantly, so only a single round trip is shown.

# [Supplementary Section 6: Other Supplementary Figures]

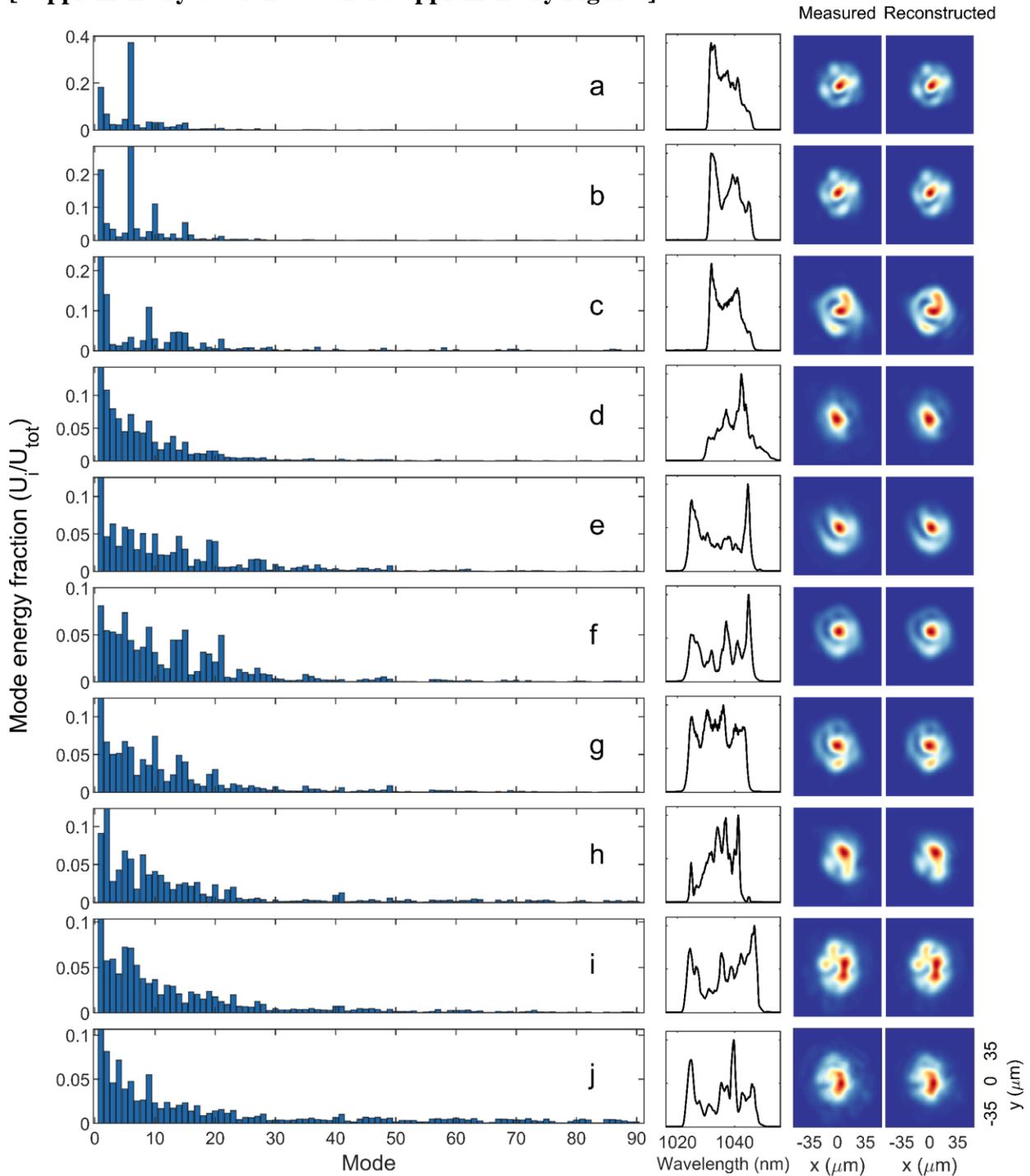

**Supplementary Figure 14: Examples of experimental STML states ranging from the very narrow filter regime (a-c) through narrow-intermediate filter sizes (d-h) and large/no spatial filter (i-j).** The panels (left to right) show the modal energy distribution, the whole-field spectrum, and the measured beam profiles. The reconstructed beam profiles obtained from the 3D field modal decomposition are shown to help qualify the mode decomposition. Overall, we see similar trends as in the few-mode simulations. In the narrow-filter regime, STML states concentrate energy into radially-symmetric modes and a small number of low-order modes. The spectra in this regime also resemble the narrow, asymmetric spectra observed in simulations. For intermediate filter sizes, we see

STML states with intermediate features, still tending to concentrate energy into radially-symmetric modes, but with a broader distribution of modes and with broader and less asymmetric, more rectangular spectra. Finally, for the largest spatial filter (here, meaning no spatial filtering besides that occurring in the optical isolator and fiber coupling) we see the experimental manifestation of the SAGE regime: broad symmetric spectra with a broad distribution of modes that is most heavily weighted to low-order modes. The absence of radial symmetric features in the beam profiles also evidences the transition from radially-symmetric low-order modes to the degenerate mode families of less-symmetric higher-order modes that occurs in the SAGE regime. Considering that 90 transverse modes are present in the oscillator, and that disordered linear mode coupling is important especially for low-symmetry modes, the agreement with the few-mode trends is significant. The experimental results also agree well with the predictions of the reduced models in Section 4. Although the neglect of Kerr nonlinearities obviously means that the spectra do not show much resemblance, the modal distributions and typical properties of the beam profiles are in good agreement.

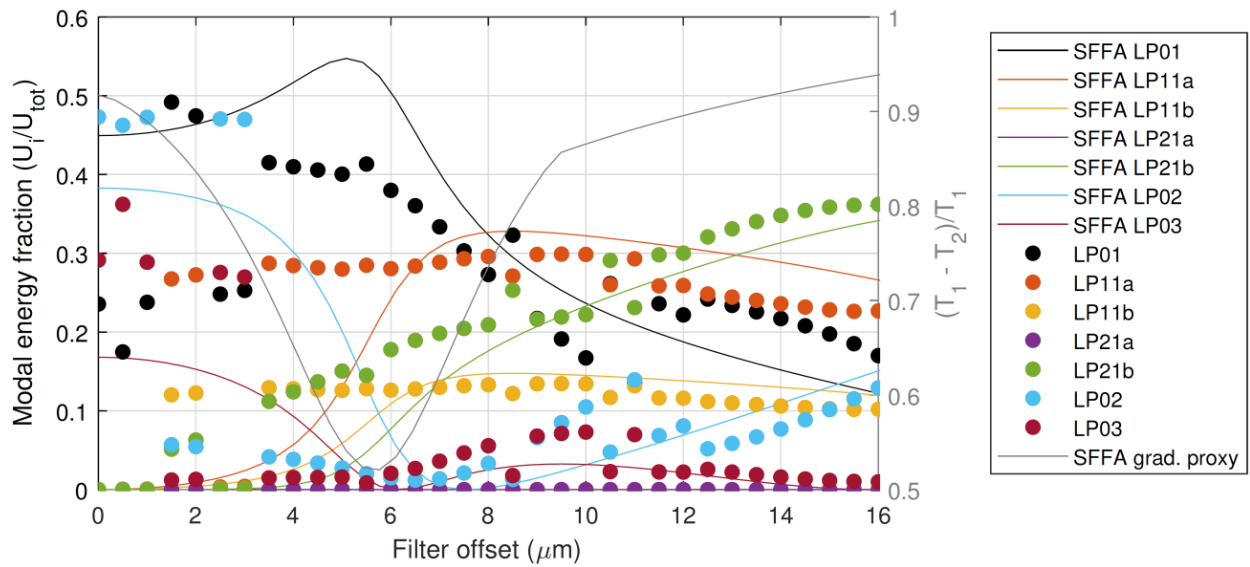

**Supplementary Figure 15: Effect of changing the spatial filter's attractor relative to the saturable absorber attractor, starting from the regime with strong competition between SFFA and SAFA.** In Fig. 2, a regime from 6 to 14 µm filter FWHM occurs where, due to the proximity of the attractors of the spatial filter and saturable absorber in their optimal pulse parameters, the eigenpulse exhibits strong pulling from each, bistability and so on. By changing the coordinates of the spatial filter field attractor (the location of its attractor in EEE) by offsetting the filter position in the transverse plane, this solution pulling is reduced and eventually the spatial filter driven regime is recovered (evident by the much closer correspondence of the SFFA and the full cavity's eigenpulse for filter offset's above about 10 µm). The plot shows the spatial filter field attractor, and the steady-state eigenpulse obtained by full simulation. The gradient proxy for the spatial filter field attractor is also shown. Here, the 12.5 µm FWHM filter is offset by the amount shown on the x-axis, in both x and y directions, relative to the center of the fiber. This plot illustrates how solution pulling depends critically on the proximity of attractors in solution space, which can be intuitively understood through the view of the EEE surface of C resulting from the addition of the independent EEE surfaces of each component.

## FROG of spatially-filtered reference

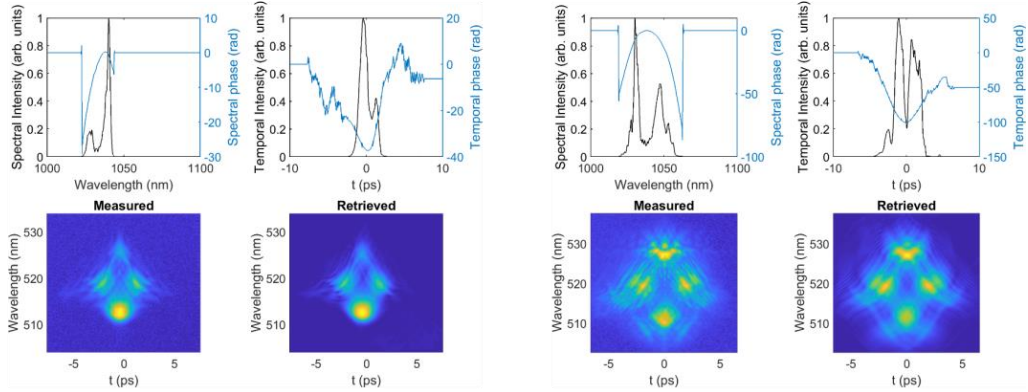

## 3D field measurement

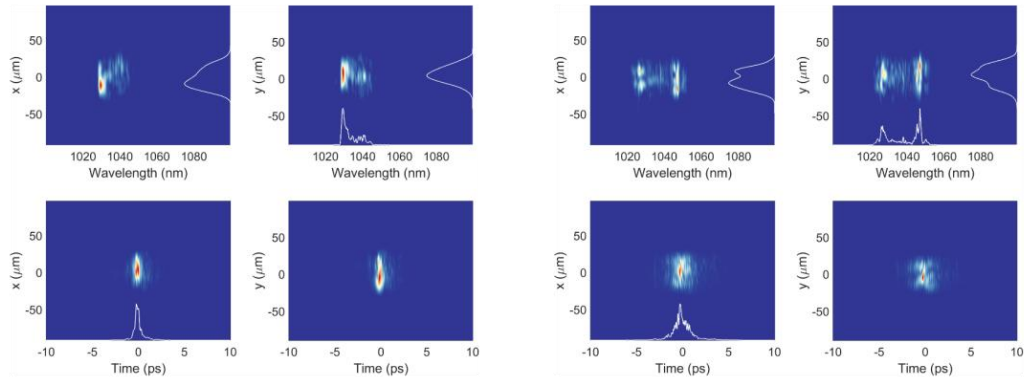

## Mode decomposition

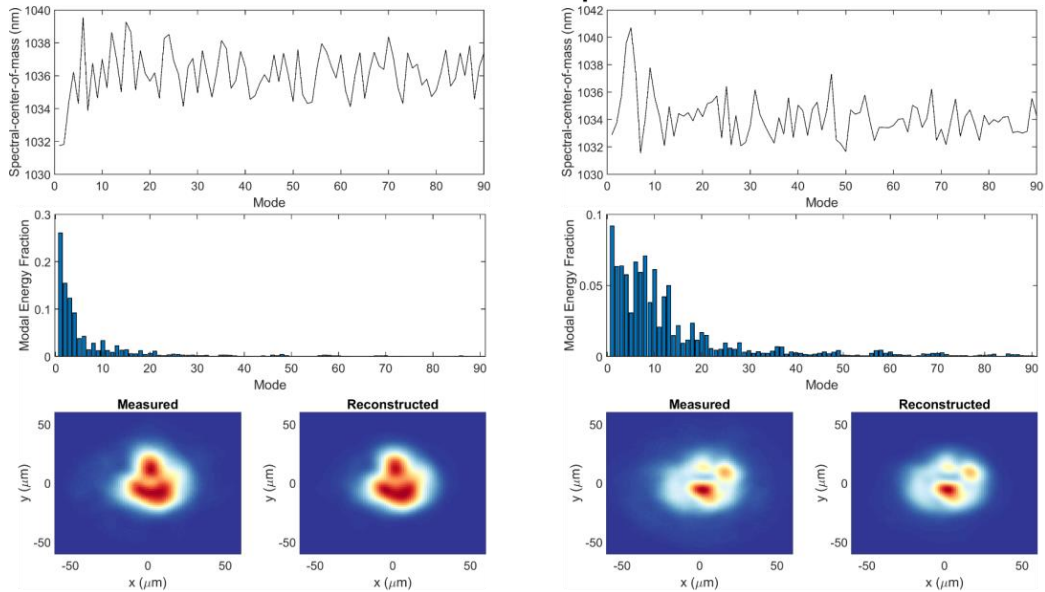

**Supplementary Figure 16: Full 3D and mode-resolved measurements for two exemplary experimental results.**
The figures show the FROG traces and retrieved fields for the spatially-filtered reference pulse, the intensity of the 3D field (in each plot, the intensity was integrated over the unplotted dimension), and the mode decomposition result, including the spectral-center-of-mass and the measured and reconstructed near-field beam profiles. The left (right) side plots correspond to a narrow (no, i.e. very large) intracavity spatial filter.

## Radio frequency spectrum

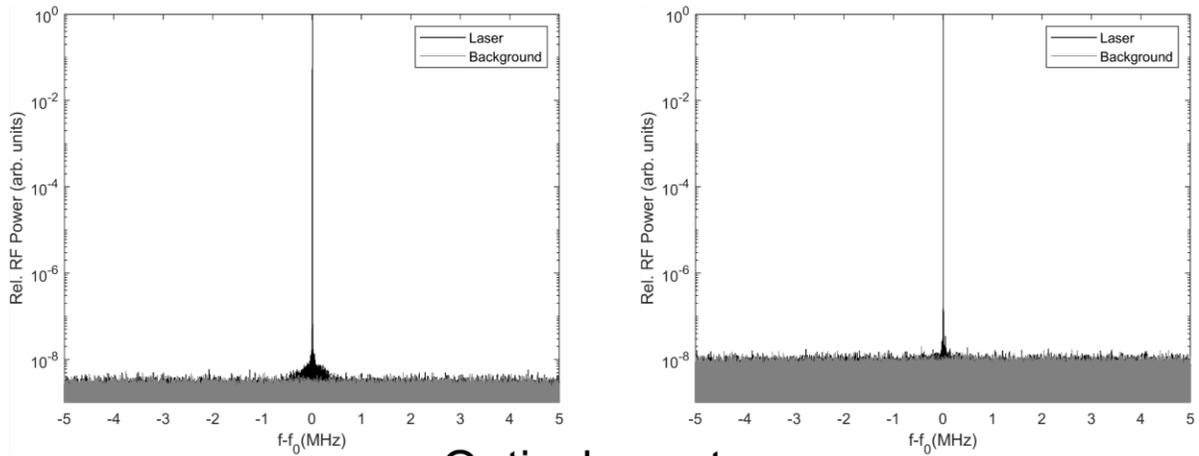

## Optical spectrum

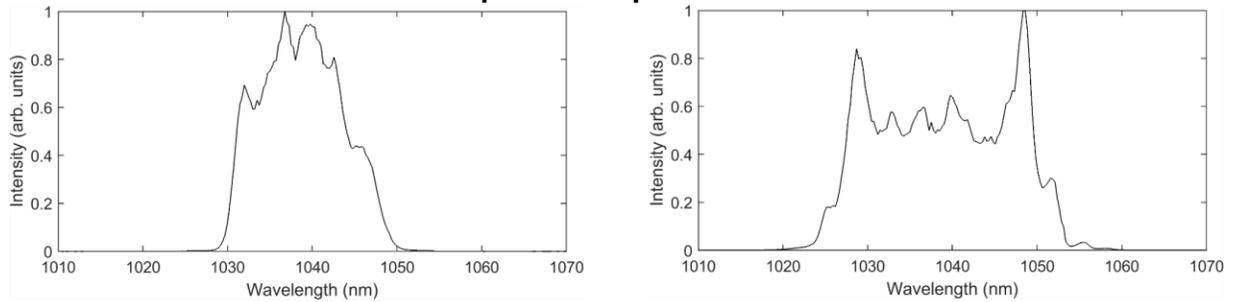

## Long-range autocorrelation

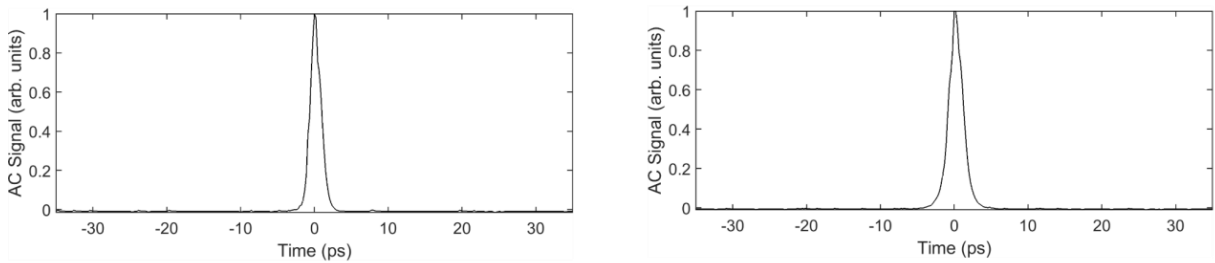

## Fast (<40 ps) photodiode measurements

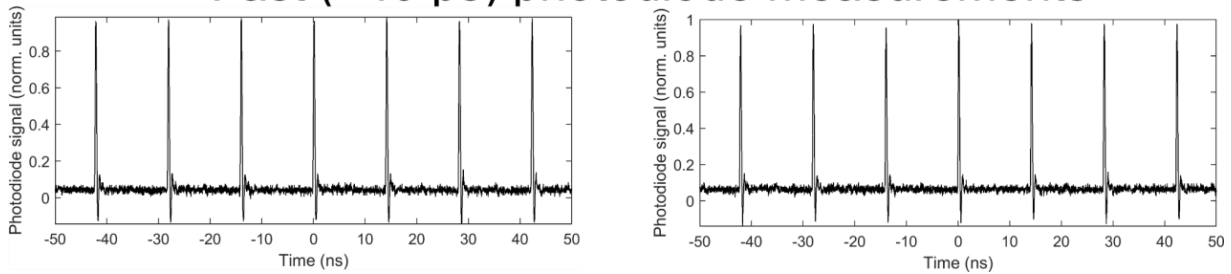

**Supplementary Figure 17: Full 1D measurements for two exemplary STML states.** The figures show the radio frequency spectrum, the optical spectrum, the long-range autocorrelation, and measurement of the pulse train using a fast photodiode and oscilloscope. Similar to the previous figure, the left (right) side plots correspond to the narrow (no) spatial filter example. For additional details, demonstration of self-starting behavior, and a tour of the experiment, see the Supplementary Movie.

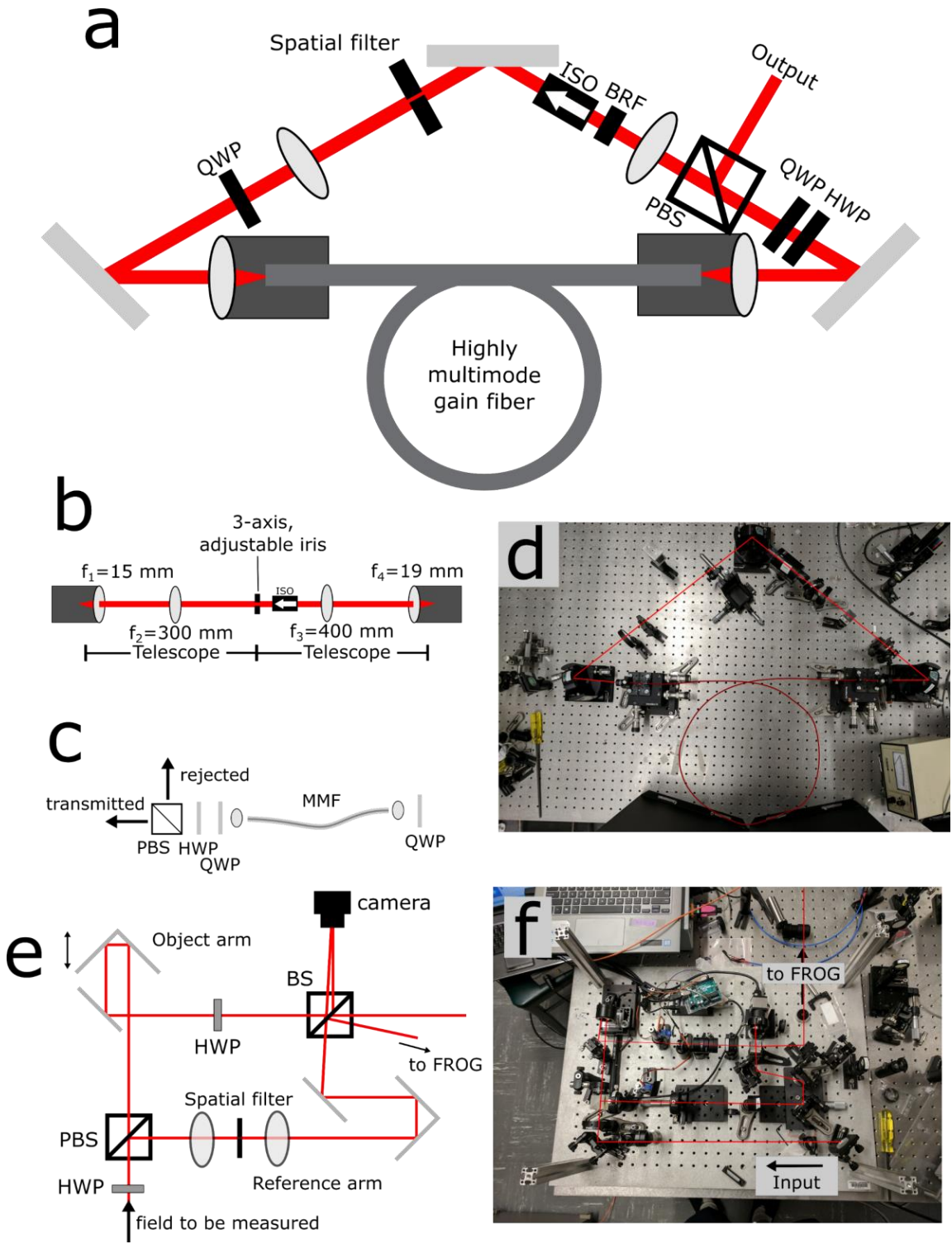

**Supplementary Figure 18: Schematics and more details on the experimental setup. a** shows the schematic of the STML oscillator, **b** shows the imaging system inside the cavity, **c** shows the implementation of MM nonlinear polarization evolution for spatiotemporal saturable absorption, **d** shows a photo of the laser cavity, with the beam path and fiber highlighted, **e** shows a schematic of the delay-scanning off-axis digital holography setup used to measure the 3D field, and **f** shows a photo of the device with beam path indicated. For more details on all, see Methods and the Supplementary Movie.

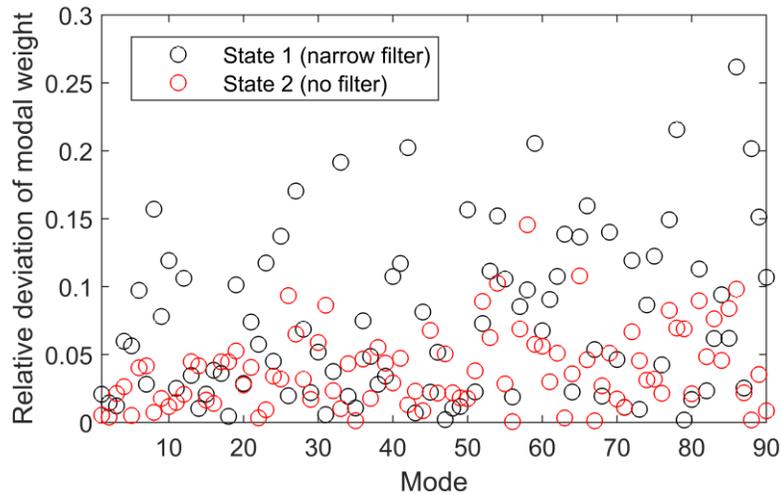

**Supplementary Figure 19: Distribution of relative uncertainty versus mode for the mode decompositions presented in the main paper.** These errors are equal to the difference (normalized to the mean modal weight) between the modal weight for a decomposition with the nominal correct image calibration and a decomposition with the fiber center position and transverse scale adjusted by the maximum deviation observed over the course of our experiments, and across multiple methods of obtaining the calibration. The mean values are 8% and 4% uncertainty for the narrow and no filter examples respectively.

### [Supplementary Section 7: Multiperiodic states]

STML oscillators also support a variety of states which exhibit more complex periods than the single-period states mainly considered so far. We mention these here because they are a common phenomenon, visible in the results we present here in one instance and noted also in Ref. [7]. While a detailed examination of these states is beyond the scope of the current work, here we give a brief summary and provide first hypotheses for the integration of these states into the theoretical framework that is the main subject of the article.

As a first relevant example, in the SAGE regime, many MM SAGE pulses exhibit multiperiodic behavior (see, e.g. the oscillating energy visible in the route-to-steady state plot shown in Supplementary Figure 12). MM SAGE pulses often display regular oscillations with a period that is not necessarily a subharmonic of the cavity repetition rate. In these pulses, no particular driving force affects the phase at a particular longitudinal position in the cavity, since the gain efficiency is maximized for any intermode phase. Consequently, free beating between modes occurs.

These non-harmonic multiperiodic states should be contrasted with other STML states that we observe: subharmonic oscillations, which correspond to higher-order eigenpulses of the cavity operator, such that $\hat{C}^l E_i(x,y,t) = E_i(x,y,t)$ for $l$ an integer. Finally, we note that noise-like, or partially-mode-locked, pulses are also easily observed in both simulation and experiment.

These more complex states could be, hypothetically (i.e., more work is needed), integrated into our theoretical framework as follows. Multiperiodic states, and noise-like pulses, represent situations where, with respect to optimizing EEE:

a) Through some form of symmetry in the cavity operator, the optimum of EEE does not depend on one or more parameters of the steady-state field, even though certain intracavity components, like the saturable absorber, are optimized for a coherent pulse. An example occurs in the previously-mentioned MM SAGE regime. In this case, a coherent mode-locked pulse forms, and the unconstrained parameters freely oscillate due to mode beating (non-harmonic multiperiodic states).
b) The EEE surface, being comprised of effects with different attractors, may simply be maximized better by a subharmonic state than a non-harmonic one. As an example, for an offset spatial filter, a certain phased combination of modes will minimize the filter loss. However, simultaneously, a different set of modal coefficients may optimize the gain efficiency, or saturable absorber. A suitable spatiotemporal evolution may satisfy both by extending the period of its self-consistency.
c) The laser simply cannot satisfy the constraints imposed by various effects to achieve mode-locking but can nonetheless approach mode-locking. This results in a noise-like pulse: a nearly-mode-locked pulse mode that might be thought of as *frustrated mode-locking*.

Supplementary Figure 20 illustrates a typical experimental observation of these multiperiodic states. Starting from an initially mode-locked state, we find that multiperiodic states (both subharmonic and not) can usually be observed by adjusting adiabatically one or more parameters of the cavity, such as the pump power or the spatial filter location.

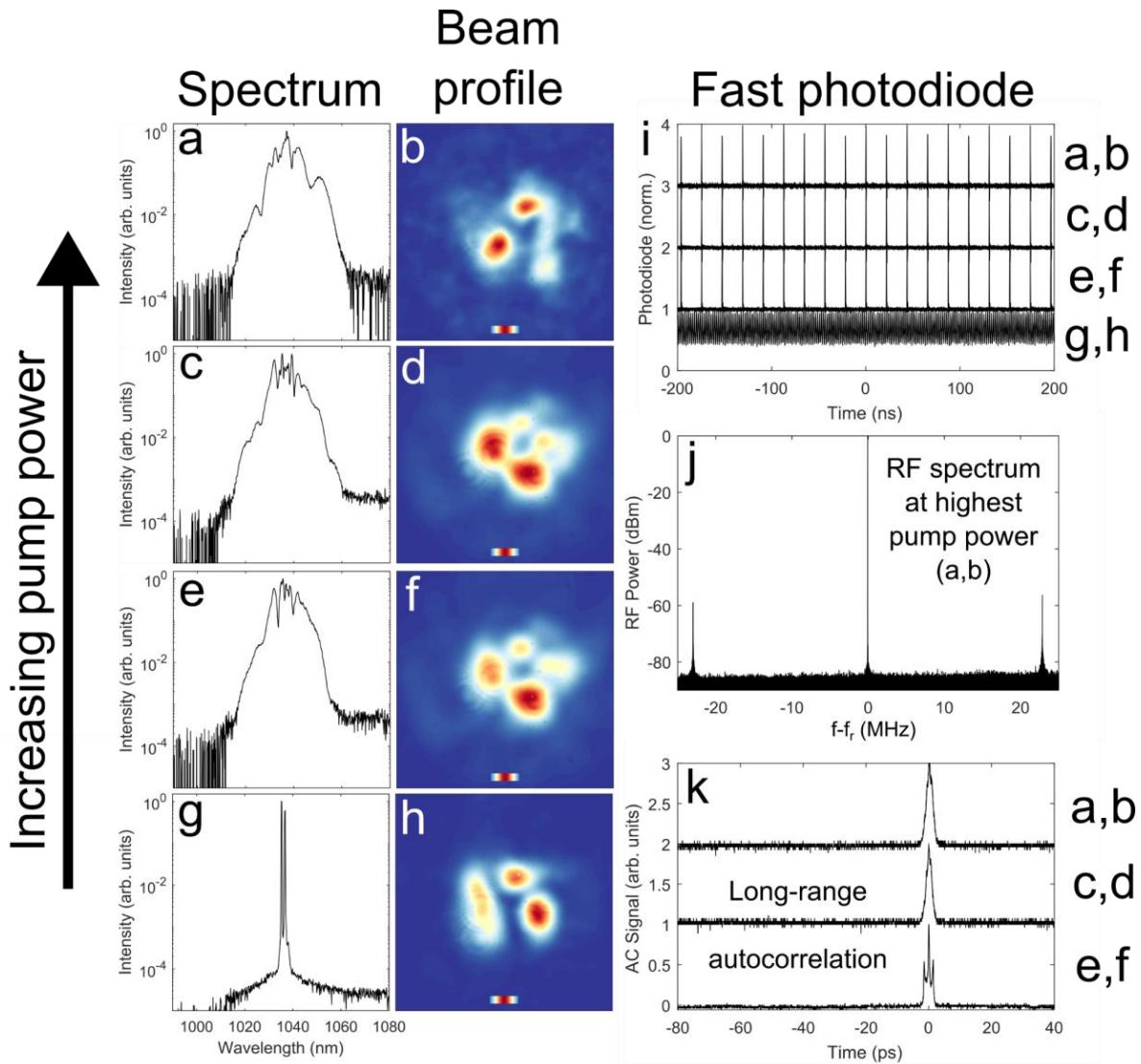

**Supplementary Figure 20: Example experimental results illustrating different regimes of STML observed when varying a single parameter (the pump power here), including subharmonic mode-locking.** At low pump power, the field operates in multiple transverse modes and a small number of longitudinal modes. As the pump power is increased, a pulse with a multi-peaked autocorrelation emerges (k). As the pump power is increased further, a single pulse is visible on the autocorrelation (k), while the beam profile and spectrum (d,c) exhibit minor changes. Finally, at the highest pump power, the beam profile exhibits a significant change (b) while the pulse train measured on a fast photodiode shows subharmonic mode-locking. The subharmonic mode-locking is also seen in the RF spectrum shown in j.

**[Supplementary Section 8: STML design guidelines and calculation of step-index few-mode fiber for STML]**

**Guidelines for STML: an overview**

For STML to be observed in general, one requires:

1. Multiple transverse modes to be guided and amplified in the laser cavity
2. A spatiotemporal saturable absorber which couples both longitudinal and transverse modes
3. Phase-sensitive nonlinear interactions (e.g., Kerr nonlinearity) between transverse and longitudinal modes
4. By some mechanism or mechanisms, net balance of group velocity and modal dispersion to establish an eigenpulse/dissipative soliton.

In the situations we have considered so far, these requirements usually mean that the group velocity dispersion should be larger than, or at most comparable to, the modal dispersion. When the dispersions meet this requirement, the broadening of pulses in each transverse mode is sufficient to ensure that, even though the pulses in each transverse mode walk-off from one another due to modal dispersion, they may still interact strongly through nonlinear interactions through the spatiotemporally-local nonlinear effects that allow STML to occur. If modal dispersion is much larger than the group velocity dispersion, then pulses in each transverse mode quickly separate in time and have little temporal overlap, thus making coupling virtually impossible. In this case, the presence of higher-order transverse modes tends to lead to a reduction in the single-pulse energy, or to instabilities of the pulse [17].

We can quantify this condition as follows. For a steady-state mode-locked pulse with a total bandwidth of $\Delta_f$, the broadening of the pulse in each mode due to group velocity dispersion is $\Delta_f L \overline{\beta_2}$, where $L$ is the cavity length, and $\overline{\beta_2}$ is the average group velocity dispersion across the cavity, and across the transverse modes of interest. If the average maximum walk-off across the pulse and over the cavity's length is $\delta\beta_1|_{\max}$, then the maximum net modal walk-off is $\delta\beta_1|_{\max} L$. The STML condition is then:

$$\delta\beta_1|_{\max} L \leq 2\Delta_f L \overline{\beta_2} \qquad (S18)$$

where the factor of 2 arises because the group velocity dispersion broadens pulses in each mode.

Of course, this inequality is meant as a rule of thumb. One might in general have a group velocity dispersion map in the cavity (e.g., by use of media with different dispersions, or geometric dispersive elements like grating pairs), or a modal dispersion map [18–20], if multiple media supporting different modes are used, or modally-dispersive elements are employed. We are also assuming normal group velocity dispersion and normal modal dispersion (meaning $\overline{\beta_2} > 0$ and, $\delta\beta_1 \geq 0$ for all modes). Although we expect that the qualitative nature of STML for different signs of either dispersion may be quite different (e.g., in the case of anomalous $\overline{\beta_2}$ and either sign of $\delta\beta_1$, one expects MM solitons similar to Refs. [21–23]), the condition mentioned above Eqn. S18 is probably still a reasonable guideline for any combination of dispersion signs. The condition would also be significantly affected by the presence of linear coupling among modes, induced by perturbation in the fiber, or by an intracavity element like a spatial filter or phase plate. In this scenario, energy among transverse modes is redistributed within the cavity as a part of the linear cavity action $\hat{T}$. If coupling of this kind is significant, the assumption of separability between the transverse and longitudinal components of the cavity eigenmodes is clearly violated. Intuitively, the coupling means that even if two modes walk-off quickly, their continuous mixing will act to prevent the breakup of the multimode pulse into multiple pulses. In the case of fiber lasers, use of other mode-like entities, like cavity or fiber principal modes, might be most appropriate in this case. Finally, in scenarios with many modes, Eqn. S18 will clearly overestimate the required bandwidth for STML to occur, since if the energy is distributed over many modes with group velocities between the minimum and maximum, modes will still efficiently nonlinearly couple to one another. A better estimate in this case might be to take $\delta\beta_1|_{max}$ as the largest walk-off rate between any two significantly-occupied modes.

As a reference point, for the 7 low-order modes of the experimental fiber considered for most of our simulations, $\delta\beta_1|_{max}$ is 0.4635 ps/m, while $\overline{\beta_2}$ is 19 fs$^2$/mm, so Eqn. S18 is satisfied for pulses with bandwidth exceeding 12 THz, or about 40 nm at 1030 nm. These bandwidths are comparable to what we observe in simulations and experiments.

**Application to step-index few-mode fiber**
In step-index fibers with a few modes, and at around 1030-nm, modal dispersion is typically large compared to the group velocity dispersion. The walk-off due to modal dispersion in typical few-mode step-index Yb-doped fibers (sold often as 'large mode area' fibers, since when coiled tightly higher-order modes are attenuated) is 1-10 ps/m. The group velocity dispersion for most modes in these fibers is close to the material dispersion, so even for a broadband pulse, the chromatic pulse broadening is much smaller than the modal walk-off. Assuming a typical value of $\overline{\beta_2}$ at 1030 nm to be 20 fs$^2$/mm in fused silica fibers with minimal waveguide dispersion, for the STML rule-of-thumb condition to be satisfied, one requires a significant pulse bandwidth of 50-500 THz (50 THz corresponds to about 180 nm at 1030 nm). 180 nm is at the extreme edge of results in single-mode ultrafast fiber lasers [24–27].

The few-mode step-index design is far from optimal for STML but can be easily tweaked to support it readily. Many design improvements (both obvious, as well as subtler) could be made by employing multiple layers, multiple coupled cores [28–30], manipulation of the fiber birefringence, or graded-index sections to the design. An optimal fiber design could improve the single-pulse energy, robustness, and ease of control of the operating modes. However, most manufacturing processes are optimized for step-index fibers. Therefore, while a detailed study of fiber design for STML is a worthwhile future study, for present purposes we consider strictly step-index fiber designs. As we show below, by proper selection of the step-index fiber parameters, designs supporting a small number of guided transverse modes, that still satisfy the STML condition for modest bandwidths, appear possible.

Supplementary Figure 21 shows how the magnitude of $\delta\beta_1$ and the required bandwidth varies for the transverse modes in step-index fibers with identical V-numbers but varying radius. The V-number is $V = NA \cdot k \cdot a$, where $NA$ is the numerical aperture of the fiber, $k = 2\pi/\lambda$ is the vacuum wavenumber, and $a$ is the radius of the step-index fiber. We choose $V = 5.5$ to concentrate on few-mode fibers, for which the study of STML would be experimentally simplified compared to many-mode fibers. In this case, we find that typically the lowest-order 6 (12, counting polarization) are sufficiently localized to the core to be considered guided. We consider the others to probably be neglectable in practice due to their high propagation loss in real fibers.

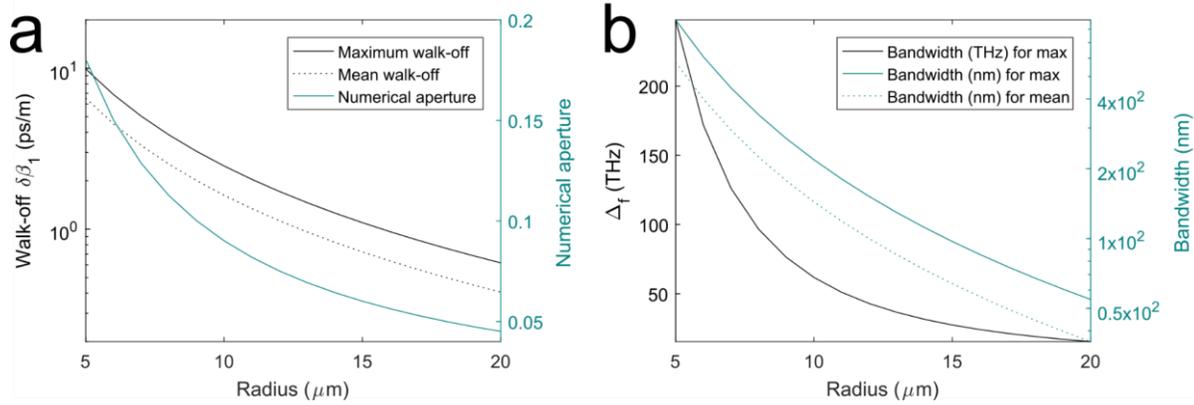

**Supplementary Figure 21: Variation of relevant parameters in a step-index few-mode fiber for fixed V-number, but varying radius. a** shows the walk-off parameters (the maximum of the 6 LP modes, and the mean across the 6 LP modes), and the numerical aperture of the fiber. **b** shows the rule-of-thumb bandwidth in frequency and wavelength units. As an additional reference, the rule-of-thumb bandwidth is shown, applied for the mean walk-off.

As an example of STML in a few-mode step-index fiber, Supplementary Figure 22 shows a cavity based on a 20 µm radius step-index fiber as above, operating in the SAGE regime (spatial filter size 40 µm FWHM). All other parameters are similar to the cavity considered in the main text, except the spectral filter bandwidth is 4 nm. The final single-pulse energy is 487 nJ.

This value should be compared to what is possible in a single-mode oscillator with a comparable effective area fiber mode. When simulations are performed with the same cavity including just the fundamental mode (which is, for this fiber, the mode with the largest area), the maximum pulse energy for which single pulsed operation is observed is 61 nJ, while the maximum single pulse energy (with multiple pulses coexisting) is 300 nJ. In other words, despite having a smaller average effective area, when multiple modes of this fiber are considered, the laser's maximum single-pulse energy is significantly improved over a single mode oscillator (approximately 10-fold here). This enhancement of laser performance by multimode operation is considered in the next section in more detail.

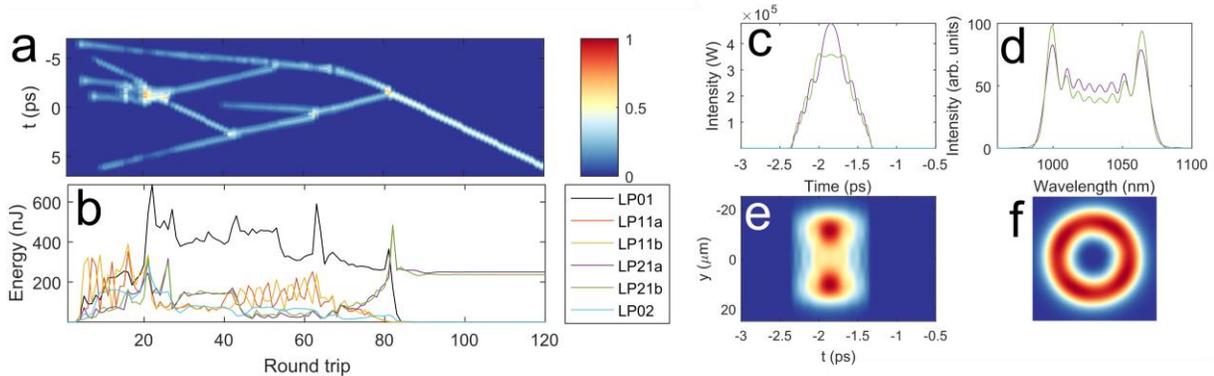

**Supplementary Figure 22: Example result showing SM SAGE mode-locking resulting from strong interpulse interactions in a few-mode step-index fiber. a** and **b** show the evolution of the field (the space-integrated temporal intensity a, and the energy in each transverse mode b) toward the steady-state, which is shown in **c-f** (the temporal and spectral intensities, mode-resolved, the spatiotemporal pulse intensity integrated over x, and the beam profile integrated over t, respectively).

### Guidelines for design of high single-pulse energy STML few-mode fiber oscillators

To investigate STML oscillators in few-mode fiber compared to single-mode fiber, we chose to study a simplified, artificial scenario that would allow many simulations to be performed. We considered the three lowest-order radially-symmetric modes of a step-index fiber with a 0.143 NA and 35 µm radius. This fiber was chosen as one whose spatial mode structure is typical (in contrast to the somewhat unusual multi-layered design of the fiber we used in much of this article).

As simulating the laser's stabilization to steady-state typically takes 1000 or more round trips, particularly when multiple pulses collide prior to the final steady-state, the fiber length was set to 25 cm, and the group velocity dispersion for all modes was set to 100 $fs^2$/mm, in order to provide similar dispersion to a typical fiber cavity despite the short length. The walk-off for LP02 was 0.56 ps/m, and for LP03 1.55 ps/m. Thus, the first STML condition is satisfied for pulse bandwidths larger than 7.75 THz (27 nm at 1030 nm). Thus, the vast majority of pulses obtained in our simulations are well within the regime where STML is expected to occur easily (and accordingly, subsequent conclusions should be interpreted in this context; outside of this regime different conclusions are drawn [17]).

While it should also be emphasized that this scenario is artificial, both in the restricted number of modes and the artificially-increased group velocity dispersion, these simulations support several conclusions that are intuitively plausible and represent a reasonable starting point for the design of future, high-power few-mode STML oscillators. We briefly summarize these design conclusions, and a few representative supporting results from our simulations here.

STML oscillators based on a relatively small number of modes may provide a net advantage in terms of single-pulse energy over single-transverse mode oscillators for several key reasons. These are:

1. The effective area of the fibers used may be much larger than a single mode fiber, by an amount that is roughly proportional to the number of guided modes (since the mode area in a step index fiber, as well as the number of modes, $M \approx V^2/2$, are both proportional to $a^2$).
2. Pulses within STML oscillators may experience a stronger linear broadening than pulses in single-transverse mode oscillators, since they have both modal and group velocity dispersion.
3. Pulses within STML oscillators may annihilate and absorb one another, since they have different EEE, and since they have different velocities permitting them to collide.

It should be noted that many other interesting advantages may be possible in specific fiber designs, or especially in the limit of many modes. However, for the present purposes, few-mode STML oscillators with conventional mode structures, these are not considered. In addition, our simulations neglect several physical effects, such as stimulated Raman scattering, linear mode coupling, and end facet damage. While we expect that design choices will allow STML oscillators to achieve super-single-mode performance even with these effects (e.g., by fiber design, spectral-filter-based artificial saturable absorbers, and end caps), but nonetheless these effects will need to be accounted for.

The advantages discussed above can easily be taken into account in a laser design by ensuring that pulses may experience strong total spatiotemporal pulse broadening and/or that pulses may efficiently collide and annihilate each other, and that the fiber used has the largest possible mode area. To permit strong spatiotemporal broadening, we choose to use a narrow spectral filter of 4 nm, as well as a strong spatiotemporal saturable absorber with a modulation depth of 100% (while this is unrealistic as has been noted earlier, it corresponds somewhat closely to nonlinear polarization rotation, and is possible with other saturable absorbing mechanisms like the Mamyshev mechanism [31–34] ). Compensating strong spatiotemporal pulse broadening by use of a narrow spatial filter is intuitively a good way to maximize pulse energy according to the conditions above. In our simulations, however, we find that the SAGE regime is typically the regime in which the highest single pulse energy is obtained, due to the multi-pulse absorption effect. SM SAGE pulses are also readily dechirped without spatiotemporal considerations, and so are the most promising form of STML pulses for competing directly with conventional sources.

Supplementary Figure 23 summarizes several sequences of simulations based on these initial parameters. While other conclusions may be tentatively drawn, the clearest conclusion is that, *for an oscillator for which STML readily occurs* (i.e., approximately within the bounds of modal-group velocity dispersion discussed earlier in this section), *modal dispersion confers a significant boost to the stable single pulse energy.*

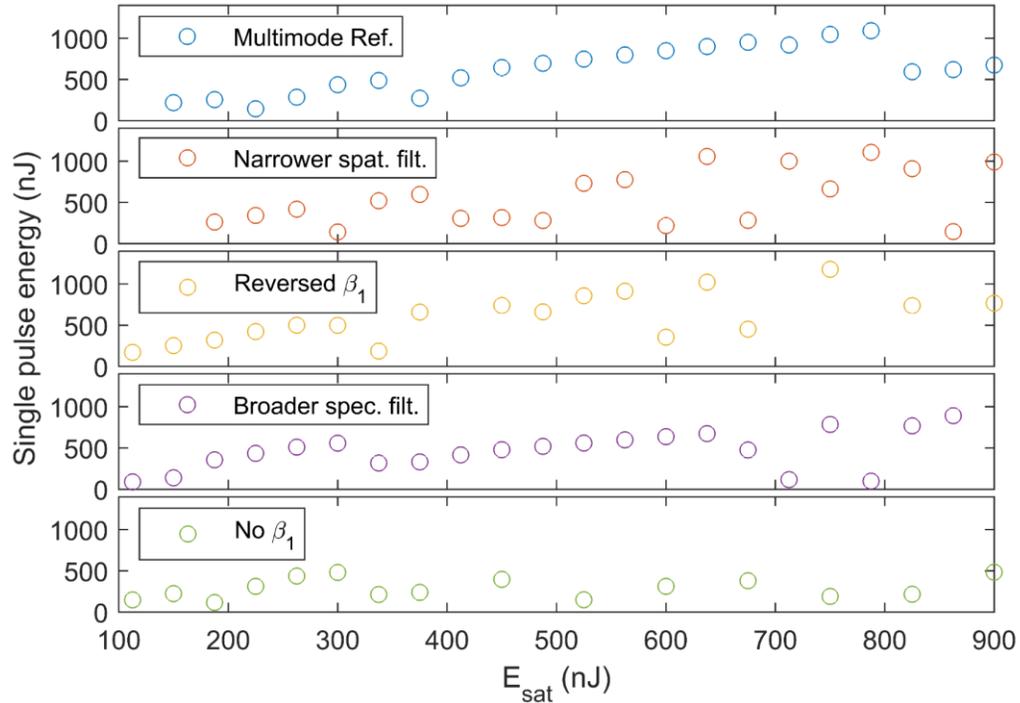

**Supplementary Figure 23: Single pulse energy obtained in various simulations of an artificial few-mode fiber laser.** The top plot show the single-pulse energy obtained with a cavity that has a 4-nm spectral and 25-µm spatial filter, other parameters noted above. The subsequent plots show the effect of reducing the spatial filter size, reversing the sign of modal dispersion, broadening the spectral filter, and turning off the modal dispersion. The most important trend easily observed is that removing the modal dispersion significantly reduces the maximum single pulse energy. Note in some instances multiple identical pulses form at the final steady-state. Here, we obtain the y-axis value as the energy per pulse. In all plots, the highest single pulse energy corresponds to a single-pulsing state.

Since the benefit of STML for allowing larger fiber mode areas is straightforward, it is worthwhile to compare the single-pulse limit of the oscillator just described to an identical oscillator with a single transverse mode whose area is equal to the fundamental mode of the STML oscillator. In a step-index fiber, this mode has the largest area of all guided modes. Supplementary Figure 25 shows the single-pulse energy in the STML and SM oscillators for increasing saturation energy in the fiber. Since the fundamental mode in the step-index fiber is the largest mode supported, this means that this imaginary single-mode oscillator has, on average, a larger effective area in this comparison. Thus, the benefit of STML must arise solely from causes 2 and 3 above. Supplementary Figure 25 shows the transverse mode composition of the multimode steady-state pulses in Supplementary Figure 24. Pulses with super-single-mode energy take on a variety of modal compositions, but the highest energy pulse contains a substantial fraction of multiple modes. Nonetheless, for slightly lower powers the pulse obtained is a SAGE pulse primarily occupying almost exclusively the fundamental mode (which as the largest mode of the fiber, represents an ideal mode to efficiently extract the available gain energy).

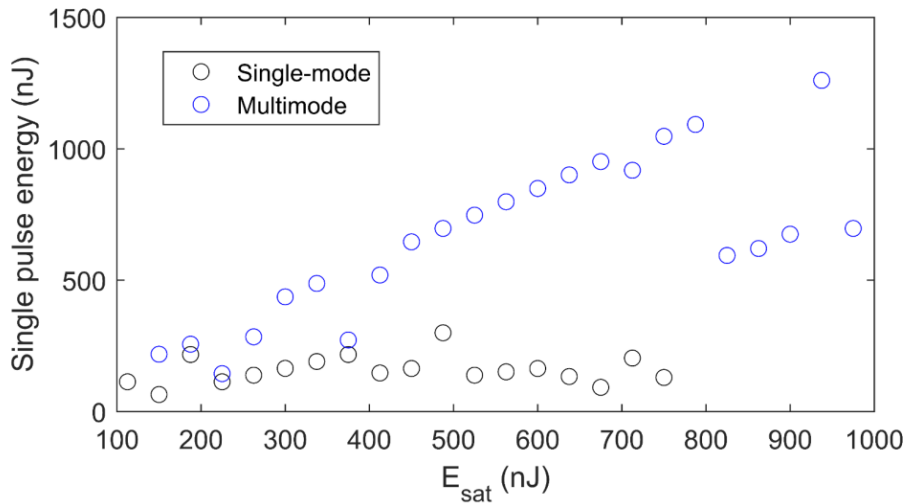

**Supplementary Figure 24: Single pulse energy obtained in a large-mode-area single-mode fiber laser compared to a multimode fiber laser with the design mentioned in the text.** The multimode oscillator produces significantly higher single-pulse energy, even though the single-mode fiber laser simulated has a mode area (artificially) equal to the largest mode in the multimode fiber laser. Note that up to 800 nJ, but especially beyond, the multimode laser does not consistently form a single pulse. Instead, for different initial noise conditions, the result of competition and collisions among multiple pulses results in a different final number of pulses. For the SM oscillator, beyond ~180 nJ saturation energy, virtually all mode-locked states are multipulsing, so the y-axis value is the energy per pulse at the steady-state. Likewise, for some points in the MM case beyond 800 nJ, the steady-state contains multiple identical pulses, and so the y-axis value is the energy per pulse. Nonetheless, our main interest in this section is in the maximum single pulse energy, *not* the maximum energy that can be obtained in a single pulse in multipulsing operation. Here, the STML oscillator here has about 7 times higher maximum single pulse energy than the SM oscillator, while the maximum energy per pulse for the saturation energies considered is just slightly less than 7.

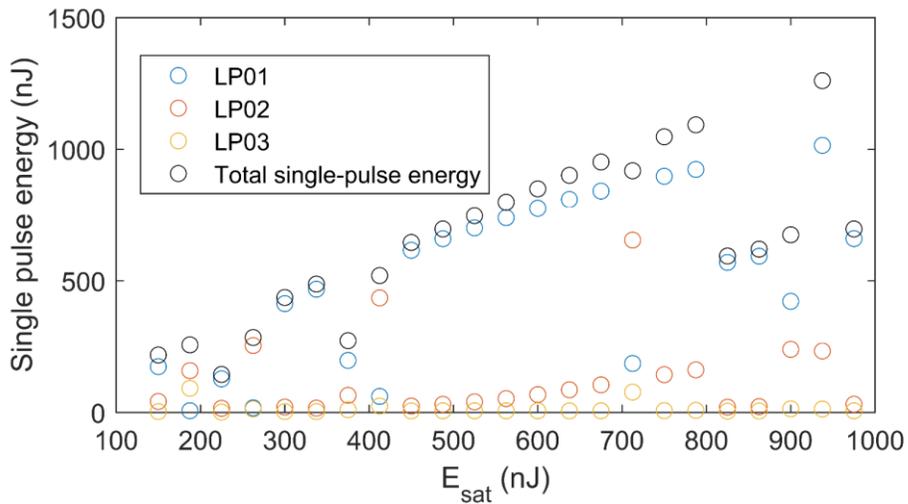

**Supplementary Figure 25: Modal composition of the steady-state pulses in the previous figure.**